\documentclass[final]{iasart_mod}           

\usepackage{graphicx}
\graphicspath{{./Fig/}}
\DeclareGraphicsExtensions{.eps}
\usepackage[utf8]{inputenc}
\usepackage[T1]{fontenc}
\usepackage{amsfonts}
\usepackage{amsmath}

\title{Morphological segmentation of hyperspectral images}
\shorttitle{Morphological segmentation of hyperspectral images}
\shortauthors{Noyel G \etal}

\author{Guillaume Noyel}
\author{Jes\'us Angulo}
\author{Dominique Jeulin}

\email{\{guillaume.noyel, jesus.angulo, dominique.jeulin\}@ensmp.fr}

\affiliation{Centre de Morphologie Math\'ematique, Ecole des Mines de
Paris, 35 rue Saint-Honor\'e, Fontainebleau, F--77305, France}

\abstract{The present paper develops a general methodology for the
morphological segmentation of hyperspectral images, i.e. with an
important number of channels. This approach, based on watershed, is
composed of a spectral classification to obtain the markers and a
vectorial gradient which gives the spatial information. Several
alternative gradients are adapted to the different hyperspectral
functions. Data reduction is performed either by Factor Analysis or
by model fitting. Image segmentation is done on different spaces:
factor space, parameters space, etc. On all these spaces the
spatial/spectral segmentation approach is applied, leading to
relevant results on the image.}

\keywords{factor analysis, hyperspectral imagery, mathematical
morphology, watershed segmentation}

\begin{document}
\begin{paper}

\section{Introduction}

Hyperspectral images are multivariate discrete functions with
typically several tens or hundreds of spectral bands. In a formal
way, each pixel of an hyperspectral image is a vector with values in
wavelength, in time, or associated with any index $j$. To each
wavelength, time or index corresponds an image in two dimensions
called channel. In the sequel we use only the term of spectrum and
spectral channel. The number of channels depends on the nature of
the specific problem under study (satellite imaging, spectroscopic
images, temporal series, etc.). Let
\begin{equation}\label{eq_im}
\mathbf{f_{\lambda}}: \left\{
\begin{array}{lll}
 E & \rightarrow & \mathcal{T}^{L}\\
 x & \rightarrow & \mathbf{f}_{\mathbf{\lambda}}(x) = \left( f_{\lambda_{1}}(x), f_{\lambda_{2}}(x), \ldots, f_{\lambda_{L}}(x) \right)\\
\end{array} \right.
\end{equation}
be an hyperspectral image, where:
\begin{description}
  \item[$\bullet$] $E \subset \mathbb{R}^{2}$, $\mathcal{T} \subset
  \mathbb{R}$ and $\mathcal{T}^{L} = \mathcal{T} \times \mathcal{T} \times \ldots \times \mathcal{T}$
  \item[$\bullet$] $x = x_{i} \ \backslash \ i\in\{1,2, \ldots, P \}$ is the spatial coordinates of a vector pixel
$\mathbf{f}_{\lambda}(x_{i})$ ($P$ is the pixels number of $E$)
  \item[$\bullet$] $f_{\lambda_{j}} \ \backslash \ j \in \{1,2, \ldots, L\}$ is a
  channel ($L$ is the channels number)
  \item[$\bullet$] $f_{\lambda_{j}}(x_{i})$ is the value of vector pixel
$\mathbf{f}_{\lambda}(x_{i})$ on channel $f_{\lambda_{j}}$.
\end{description}

In this paper, we introduce a general methodology for hyperspectral
image segmentation, using a watershed based approach. The watershed
transformation is a powerful tool for mathematical morphology
segmentation \citep{serra:1982,soille:2003}.

Watershed segmentation requires a gradient (i.e. a scalar function)
and markers on the target structures to obtain a correct image
segmentation \citep{beucher:1992}. A gradient on a multivariate
function can be obtained in different ways. One way is to calculate
on each image channel a modulus of a gradient, and to take the sum
or the supremum of the gradients \citep{meyer:1992,angulo:2003}.
Another way is to use vectorial gradients based on distance between
vector pixels \citep{angulo:2003,evans:2006}. We consider here
various alternatives for hyperspectral images.

Moreover, when dealing with hyperspectral images, the large number
of channels generates data redundancies. Consequently, it is
necessary to reduce the amount of data, to extract pertinent
information. To do this, two ways are explored: a linear factor
analysis and a model approach.

Previously in the literature, several approaches to multispectral
image segmentation were explored. \citet{flouzat:1998} use a spatial
and a spectral segmentation based on the filtering of the image
adjacency graph. \citet{paclik:2003} obtain, with statistical
classifiers, the pixels probabilities of membership to clusters for
spectral domain and the pixels probabilities of membership to
clusters for spatial domain. The pixels probabilities of membership
to a cluster are obtained by multiplying both probabilities, because
they assume independence between spatial and spectral information.
The pixels are classified and the process is repeated until
convergence. \citet{li:2004} compute on each channel a morphological
multiscale gradient by summation of morphological gradients with
increasing size structuring elements. As watershed requires a scalar
gradient (i.e. one channel), the channels gradients are summed with
a weight equal to one. After morphological filtering on the
gradient, the local minima are used as markers for watershed
segmentation. \citet{scheunders:2001} computes a gradient by
summation of channels gradients followed by a watershed
segmentation. \citet{soille:1996} combines spectral classification
on histograms and spatial segmentation. The multidimensional
histogram is segmented, using the watershed algorithm, to obtain a
classified image. On the classified image the minima of the gradient
of the hyperspectral image are imposed. With the gradient and the
markers he applies the watershed segmentation.

The methodology and the different alternatives studied in the
current paper are illustrated by means of a 60 channels image
acquired by active thermography on plastic lids. The size of the
image is $256 \times 256$ pixels $\times \ 60$ channels. The aim is
to segment glue occlusions within plastic lids. This sequence of
images comes from Laboratoire Le2i, Le Creusot, France.
\citet{legrand:2002} explain how the image was acquired. They also
present a segmentation that was done on a channel, using difference
of lid images with and without glue, thresholding and filtering.
Their segmentation is used as a reference for comparison (fig.
\ref{fig_channels_truth}). In our case, only the sequence with glue
occlusions is available. That's why we cannot use image difference
or any kind of calibration with a reference image.

\begin{figure}
\begin{center}
\begin{tabular}{cc}
    \includegraphics[width=0.4\columnwidth]{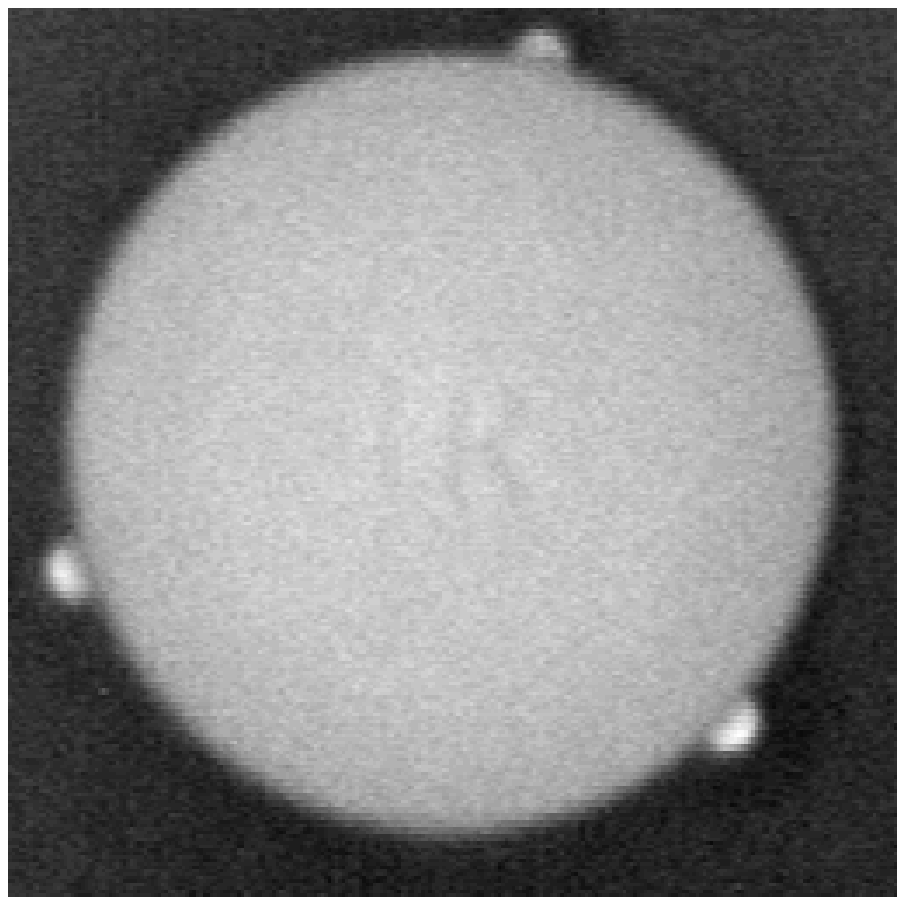}&
    \includegraphics[width=0.4\columnwidth]{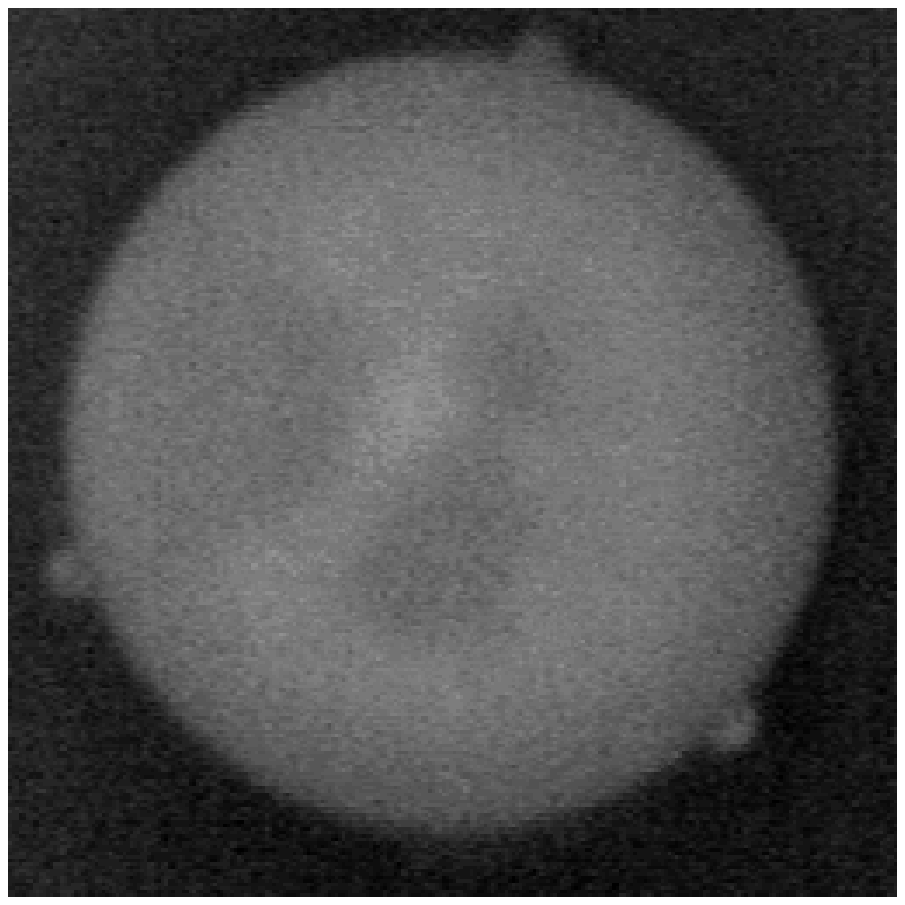}\\
    $f_{\lambda_{1}}$ & $f_{\lambda_{30}}$\\
    \includegraphics[width=0.4\columnwidth]{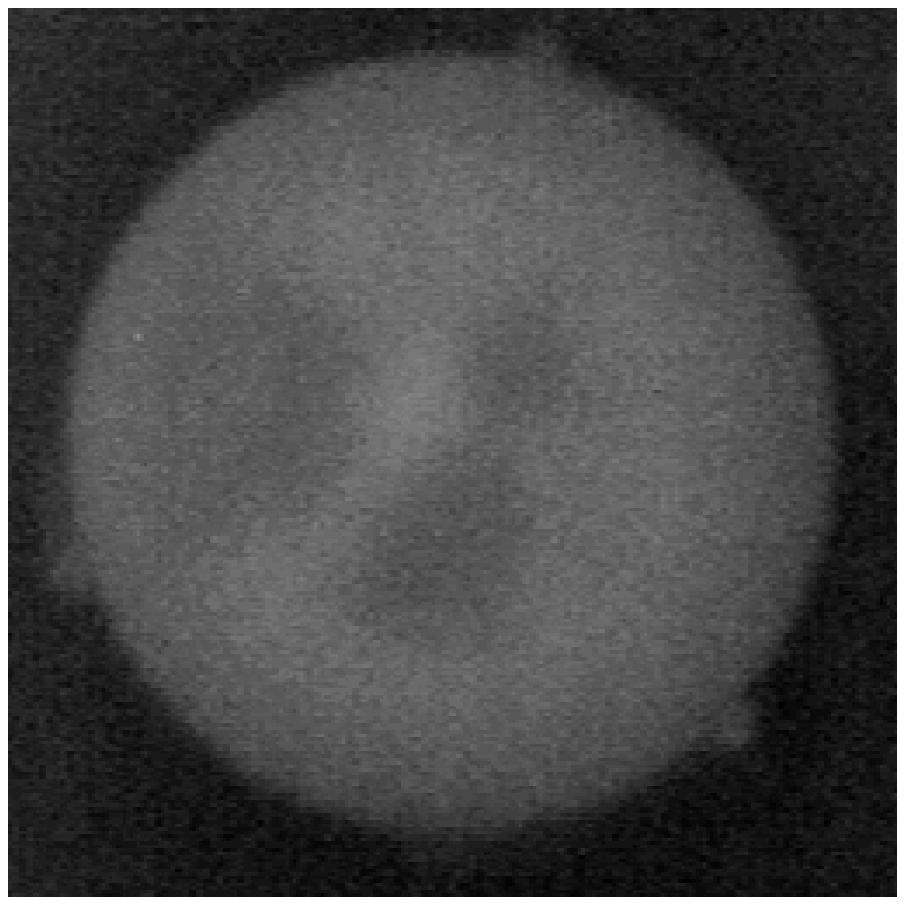}&
    \includegraphics[width=0.4\columnwidth]{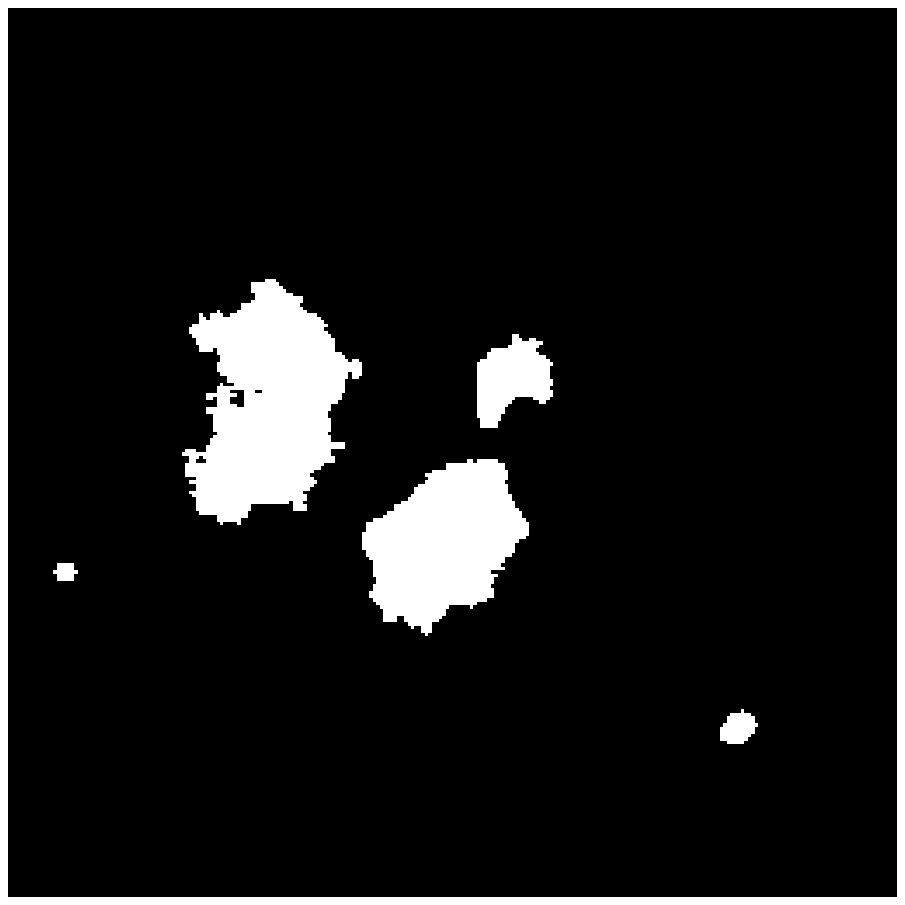}\\
    $f_{\lambda_{60}}$ & reference \\
\end{tabular}
\end{center}
\caption{Three channels of the original image and the reference
obtained by \citet{legrand:2002} method.} \label{fig_channels_truth}
\end{figure}

\section{Framework for morpho- \newline logical segmentation on \newline multivariate data}

The watershed transformation is one of the most powerful tools for
segmenting images. According to a flooding paradigm, the watershed
lines associate a catchment basin to each minimum of the function
\citep{beucher:1992}. Typically, the function to flood is a gradient
function which defines the transitions between the regions. Using
the watershed on a scalar image without any preparation leads to a
strong over-segmentation (large number of minima). There are two
alternatives in order to get rid of the over-segmentation. The first
one consists in initially determining markers for each region of
interest: using the homotopy modification, the local minima of the
gradient function are only the region markers. A difficult issue is
to determine the markers, especially for generic images. The second
alternative involves hierarchical approaches based on non-parametric
merging of catchment basins (waterfall algorithm) or based on the
selection of the most significant minima according to different
criteria (dynamics, area or volume extinction values)
\citep{meyer:2001}.

In this paper, we focus on markers based segmentation. In fact, we
consider that the hyperspectral images have enough information to
define markers from a spectral classification.

\textbf{Multivariate gradients}

A gradient image, in fact the norm, is a scalar function with values
in the reduced interval $[0, 1]$, i.e. $\nabla: E \rightarrow
[0,1]$. This normalization is always applied to multivariate
gradients given below in order to have homogeneous gradient
functions. To define a gradient, four approaches are considered: a
morphological gradient on one channel, a metric-based gradient on
all channels, a gradient defined as the supremum, or as the sum, of
morphological gradients on each channel. The morphological gradient
is a marginal gradient (i.e. it can only be applied on scalar
images) defined as the difference between the channel dilation and
erosion with a structuring element $B_x$ which is the neighborhood
of point $x \in E$ \citep{serra:1982}:
\begin{eqnarray}\label{eq_morphological_gradient}
    g(f_{\lambda_{j}}(x)) & = & \delta_{B_x}(f_{\lambda_{j}}(x)) -
    \varepsilon_{B_x}(f_{\lambda_{j}}(x)) \\ \nonumber
                          & = & \vee_{B_x}(f_{\lambda_{j}}(x)) -
                                \wedge_{B_x}(f_{\lambda_{j}}(x))
\end{eqnarray}

The gradient supremum of morphological gradients on each channel is
a vectorial gradient defined as:
\begin{equation}\label{eq_sup_gradient}
    \nabla_{\vee}\mathbf{f_{\lambda}}(x) = \vee [g(f_{\lambda_{j}}(x)), j \in \{1,\ldots,L\}]
\end{equation}
Each morphological gradient must be normalized between $[0,1]$
before taking the supremum.

The gradient weighted sum of morphological gradients is given by:
\begin{equation}\label{eq_sum_gradient}
    \nabla_{+}\mathbf{f_{\lambda}}(x) = \sum_{j=1}^{L} w_{\lambda_{j}}g(f_{\lambda_{j}}(x))
\end{equation}
where $w_{\lambda_{j}}$ denotes the weight of the gradient of
channel $f_{\lambda_{j}}$.

The metric-based gradient is a vectorial gradient defined as the
difference between the supremum and the infimum of a defined
distance on a unit neighbourhood $B(x)$:
\begin{eqnarray}\label{eq_euclidian_gradient}
    \nabla_{d}\mathbf{f_{\lambda}}(x) &=& \vee[ d(\mathbf{f_{\lambda}}(x), \mathbf{f_{\lambda}}(y)) , y \in B(x)] - {}
    \nonumber\\
    & &{} \wedge[ d(\mathbf{f_{\lambda}}(x), \mathbf{f_{\lambda}}(y)) , y \in B(x)
    ]\nonumber\\.
\end{eqnarray}
Various metric distances are available for this gradient such as:
\begin{description}
  \item[$\bullet$] Euclidean distance:
\begin{equation}\label{eq_dist_euclidean}
 d_{E}(\mathbf{f_{\lambda}}(x), \mathbf{f_{\lambda}}(y)) =
 \sqrt{ \sum_{j=1}^{L}( f_{\lambda_{j}}(x) - f_{\lambda_{j}}(y) )^2  }
\end{equation}
  \item[$\bullet$] Mahalanobis distance:
\begin{eqnarray}\label{eq_dist_mahalanobis}
  d_{M}(\mathbf{f_{\lambda}}(x), \mathbf{f_{\lambda}}(y)) &=& {}\nonumber\\
  {} \sqrt{(\mathbf{f_{\lambda}}(x) - \mathbf{f_{\lambda}}(y))^{t}
  \Sigma^{-1}(\mathbf{f_{\lambda}}(x) - \mathbf{f_{\lambda}}(y))}
\end{eqnarray}
where $\Sigma$ is the covariance matrix between variables (channels)
of $\mathbf{f_{\lambda}}$. If channels are uncorrelated, the
covariance matrix is diagonal. The diagonal values are equal to
channels variance $\sigma^{2}_{\lambda_{j}} \ \backslash \ j \in
\{1,2, \ldots, L\}$. Therefore, the Mahalanobis distance becomes:
\begin{equation}\label{eq_dist_mahalanobis_simplified}
  d_{M}(\mathbf{f_{\lambda}}(x), \mathbf{f_{\lambda}}(y)) =
   \sqrt{ \sum_{j=1}^{L} \left( \frac{f_{\lambda_{j}}(x) - f_{\lambda_{j}}(y)}{\sigma_{\lambda_{j}}} \right)^{2} }
\end{equation}
  \item[$\bullet$] chi-squared distance:
\begin{eqnarray}\label{eq_dist_chi2}
    d_{\chi^{2}}( \mathbf{f}_{\lambda}(x_{i}) ,
    \mathbf{f}_{\lambda}(x_{i'}) ) &=& {} \nonumber\\ {} \sqrt{\sum_{j=1}^{L}
    \frac{N}{f_{.\lambda_{j}}} \left( \frac{ f_{\lambda_{j}}(x_{i}) }{
    f_{x_{i}.} } - \frac{ f_{\lambda_{j}}(x_{i'}) }{ f_{x_{i'}.} }
    \right)^{2}}
\end{eqnarray}
with $f_{.\lambda_{j}} = \sum_{i=1}^{P}
f_{\lambda_{j}}(x_{i})$, $f_{x_{i}.} = \sum_{j=1}^{J}
f_{\lambda_{j}}(x_{i})$ and $N =
\sum_{j=1}^{L}\sum_{i=1}^{P}f_{\lambda_{j}}(x_{i})$.
\end{description}

\textbf{Markers by spectral clustering}

The markers defining the targets are obtained with an unsupervised
classification based on clustering. We have used in this study the
clustering algorithm "Clara" \citep{kaufman:1990}. This is a similar
way, but more robust than the "kmeans" classification, in order to
cluster large numbers of points. Then, the pertinent clusters are
selected to be the markers. To perform clustering, each channel is
considered as a variable. Therefore each pixel has its own value on
each variable.

\section{Data reduction and filtering using factor analysis}

Due to the redundancy of channels, a data reduction is performed
using Factor Correspondence Analysis (FCA) \citep{benzecri:1973}. We
prefer an FCA in place of a Principal Component Analysis because
image values are positive and the spectral channels can be
considered as probability distributions. The metric used in FCA is
the chi-squared normalized by channels weight. This metric is
adapted to probability laws. FCA can be seen as a transformation
from image space to factorial space (eq. \ref{eq_FCA}). In factorial
space the coordinates of the pixels vector on each factorial axis
are called pixels factors. The pixels factors can be considered as
an hyperspectral image whose channels correspond to factorial axes:
\begin{equation}\label{eq_FCA}
    \zeta: \left\{
\begin{array}{lll}
 \mathcal{T}^{L} & \rightarrow & \mathcal{T}^{K} \text{ / } K < L \\
 \mathbf{f}_{\mathbf{\lambda}}(x) & \rightarrow & \mathbf{c}^{\mathbf{f}}_{\alpha}(x) =
  \left( c^{\mathbf{f}}_{\alpha_{1}}(x), \ldots, c^{\mathbf{f}}_{\alpha_{K}}(x) \right) \\
\end{array} \right.
\end{equation}
A limited number $K$ of factorial axes is usually chosen. Therefore
FCA can be seen as a projection of initial pixels in a factor space
with a smaller dimension.

The pseudo-inverse transform consists in reconstructing the images
from factors (eq. \ref{eq_FCA_inverse}). This is an approximation of
the original image if one keeps a part of factorial axes for the
reconstruction:
\begin{equation}\label{eq_FCA_inverse}
    \widehat{\zeta}^{-1}: \left\{
\begin{array}{lll}
 \mathcal{T}^{K} & \rightarrow & \mathcal{T}^{L} \text{ / } K < L \\
 \mathbf{c}^{\mathbf{f}}_{\alpha}(x) & \rightarrow & \mathbf{\widehat{f}}_{\mathbf{\lambda}}(x) =
 \left( \widehat{f}_{\lambda_{1}}(x), \ldots, \widehat{f}_{\lambda_{L}}(x) \right)\\
\end{array} \right.
\end{equation}
Besides, hyperspectral images usually contain noise due to the
acquisition device, compression, etc. In this case it is possible to
use FCA to filter images. As shown by \citet{green:1988}, the noise
is rejected on the last factorial axes whereas the signal remains on
the first axes. By keeping only axes with sufficient signal and
reconstructing the image, all channels are filtered. In our example,
the image is very noisy ; consequently better results are obtained
by filtering the whole sequence with a FCA, keeping the two first
factorial axes (fig. \ref{Fig_canal_50_before_after_filtering}). The
eigenvalues associated with these two factors are $\mu_{1} = 13.9 \:
10^{-4}$ and $\mu_{2}=3.15 \: 10^{-4}$. In other words, the first
axis represent $13.8 \%$ of inertia and the second one $3.1 \%$.
Therefore, the inertia of the two first axes is equal to $16.9 \%$
which is small. In fact the next axes contain a lot of noise. This
explains why their share in inertia is high. Therefore, they must be
removed to get information without noise.

\begin{figure}
\begin{center}
\begin{tabular}{cc}
  \includegraphics[width=0.3\columnwidth]{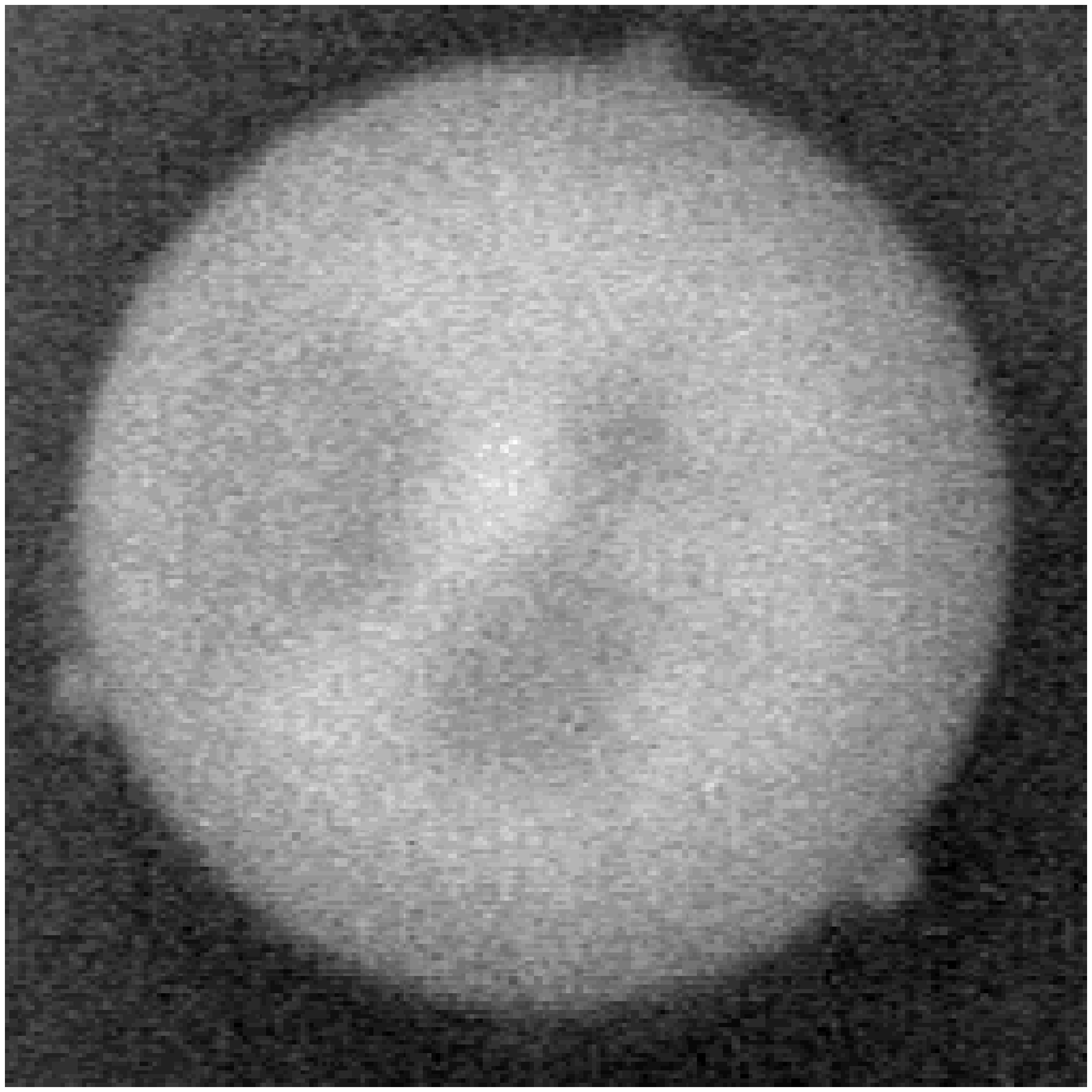}&
  \includegraphics[width=0.3\columnwidth]{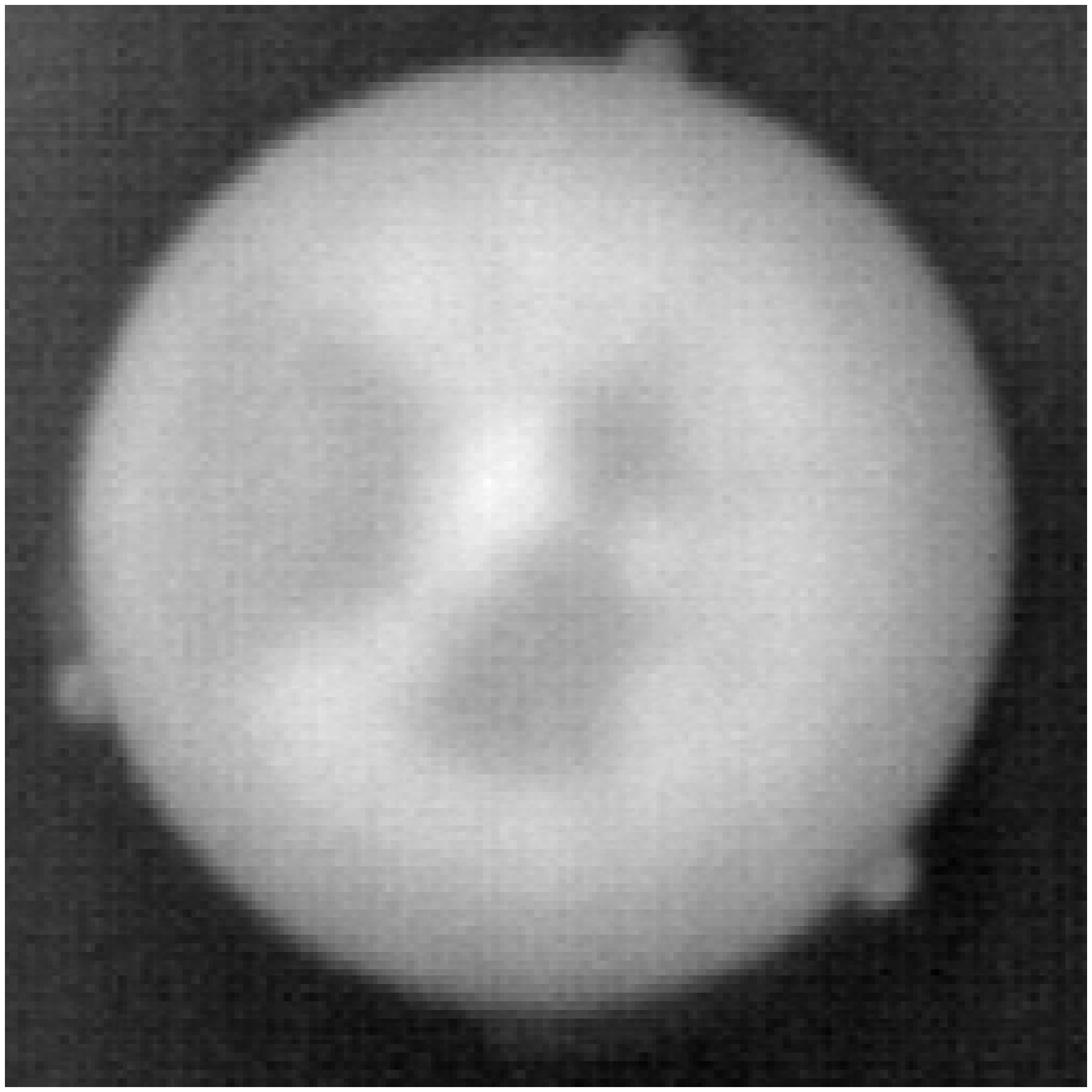} \\
  $f_{\lambda_{50}}$ & $\widehat{f}_{\lambda_{50}}$\\
  \end{tabular}
\end{center}
\caption{$f_{\lambda_{50}}$ and $\widehat{f}_{\lambda_{50}}$ are the
channel 50 before and after filtering by FCA.}
\label{Fig_canal_50_before_after_filtering}
\end{figure}

\subsection{Segmentation of
$\mathbf{\widehat{f}}_{\mathbf{\lambda}}(x)$, $
\mathbf{c}^{\mathbf{f}}_{\alpha}(x)$}

It is possible to generate a segmentation on the hyperspectral image
composed of factorial axes $\mathbf{c}^{\mathbf{f}}_{\alpha}(x)$
(fig. \ref{Fig_2_axis_FCA}) or on the filtered image
$\mathbf{\widehat{f}}_{\mathbf{\lambda}}(x)$.

\begin{figure}
\begin{center}
\begin{tabular}{cc}
  \includegraphics[width=0.3\columnwidth]{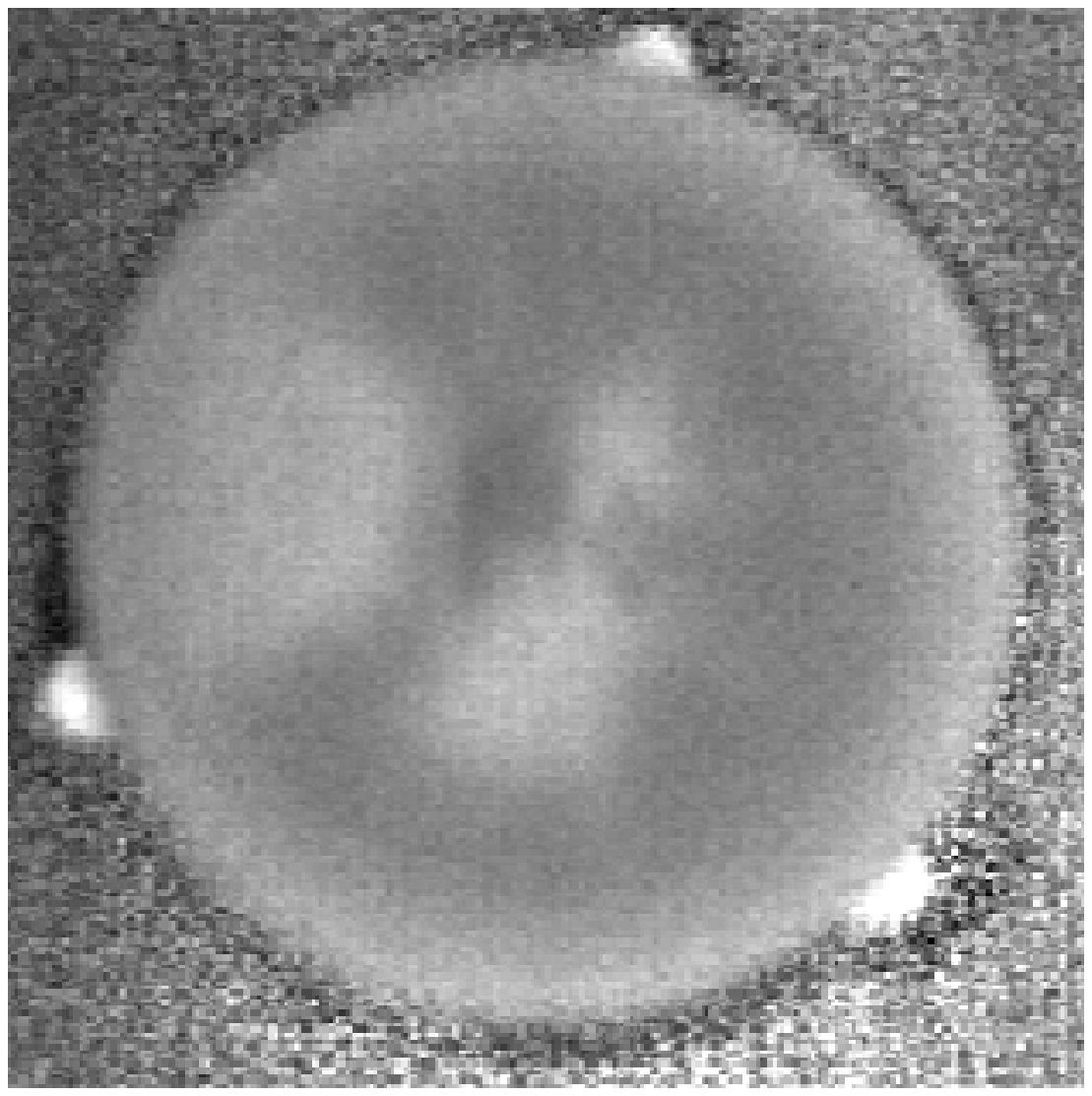}&
  \includegraphics[width=0.3\columnwidth]{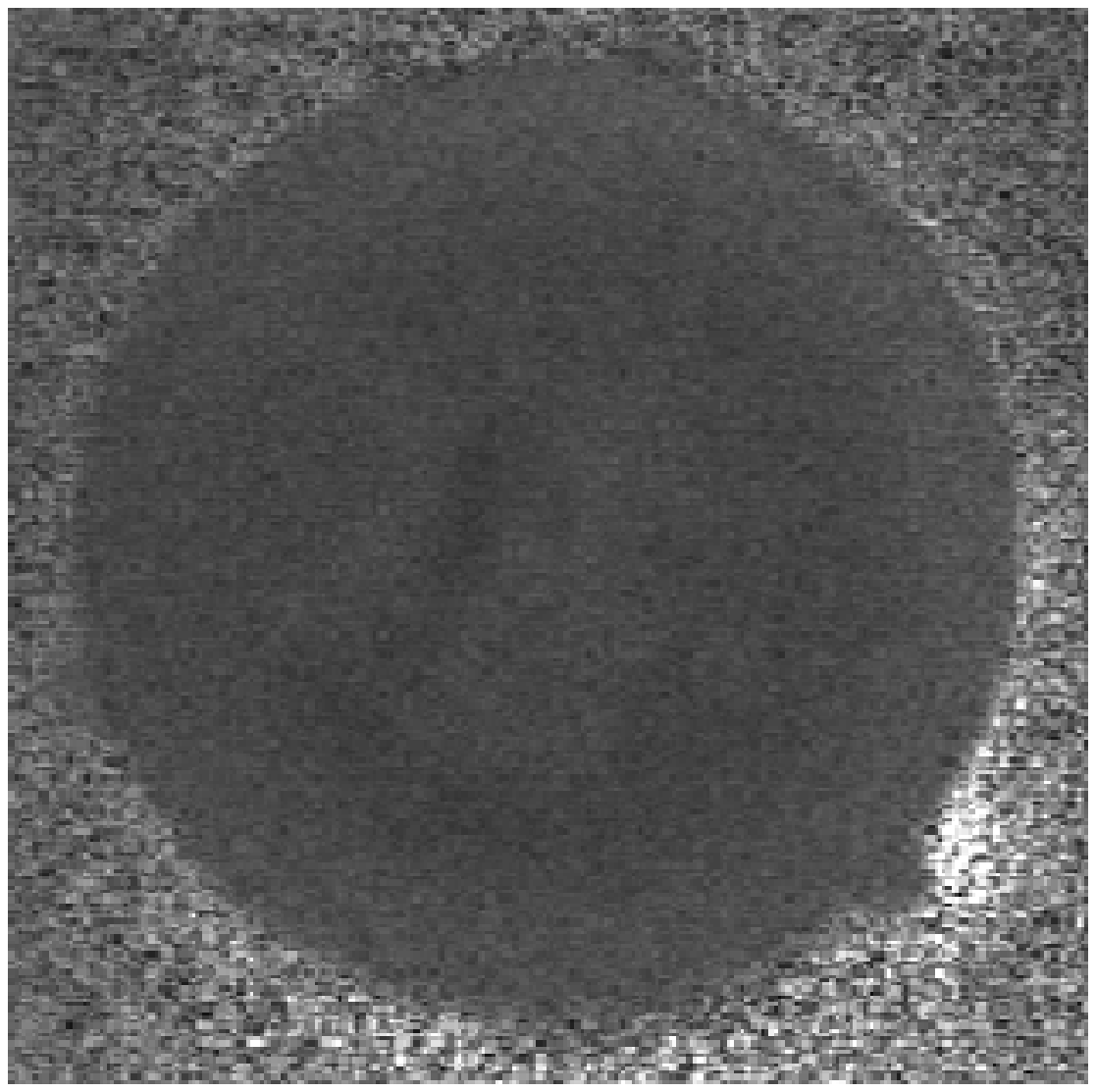} \\
  $c^{\mathbf{f}}_{\alpha_{1}}(x)$ & $c^{\mathbf{f}}_{\alpha_{2}}(x)$ \\
  \end{tabular}
\end{center}
\caption{FCA factors on axes $1$ and $2$ of original image. For
visualization, factors are scaled by a factor.}
\label{Fig_2_axis_FCA}
\end{figure}
In the factorial space made up of the two first axes, a
classification by "Clara" is processed. Then the green cluster
corresponding to glue is selected defining the markers (fig.
\ref{Fig_segmentation_axes_im}). In order to regularize these
markers, an opening with an hexagonal structuring element of size 5
is applied. As differences between glue and lid are small, a
gaussian filter of size 11 followed by a morphological leveling are
applied on each channel, to enhance the contours and to obtain a
better gradient. The levelings are a subclass of symmetric connected
filters (or filters by reconstruction) that suppress details but
preserve the contours of the remaining objects~\citep{meyer:2004}.
The levelings need an image marker, a rough simplification of the
reference image, to determine the structures to be leveled.

The Euclidean distance in FCA factorial space is equivalent to the
chi-squared distance in image space \citep{benzecri:1973}. Therefore
a chi-squared distance based gradient
$\nabla_{\chi^{2}}\mathbf{\widehat{f}_{\lambda}}(x)$ is performed on
the filtered image $\mathbf{\widehat{f}}_{\mathbf{\lambda}}(x)$ and
an Euclidean distance based gradient
$\nabla_{E}\mathbf{c}^{\mathbf{f}}_{\alpha}(x)$ on the FCA factors
$\mathbf{c}^{\mathbf{f}}_{\alpha}(x)$. Then the watershed
segmentation is computed (fig. \ref{Fig_segmentation_im} and
\ref{Fig_segmentation_axes_im}). Both segmentations, compared to the
reference model (fig. \ref{fig_channels_truth}), are not
satisfactory for the present image.

\begin{figure}
\begin{center}
\begin{tabular}{cc}
  \includegraphics[width=0.3\columnwidth]{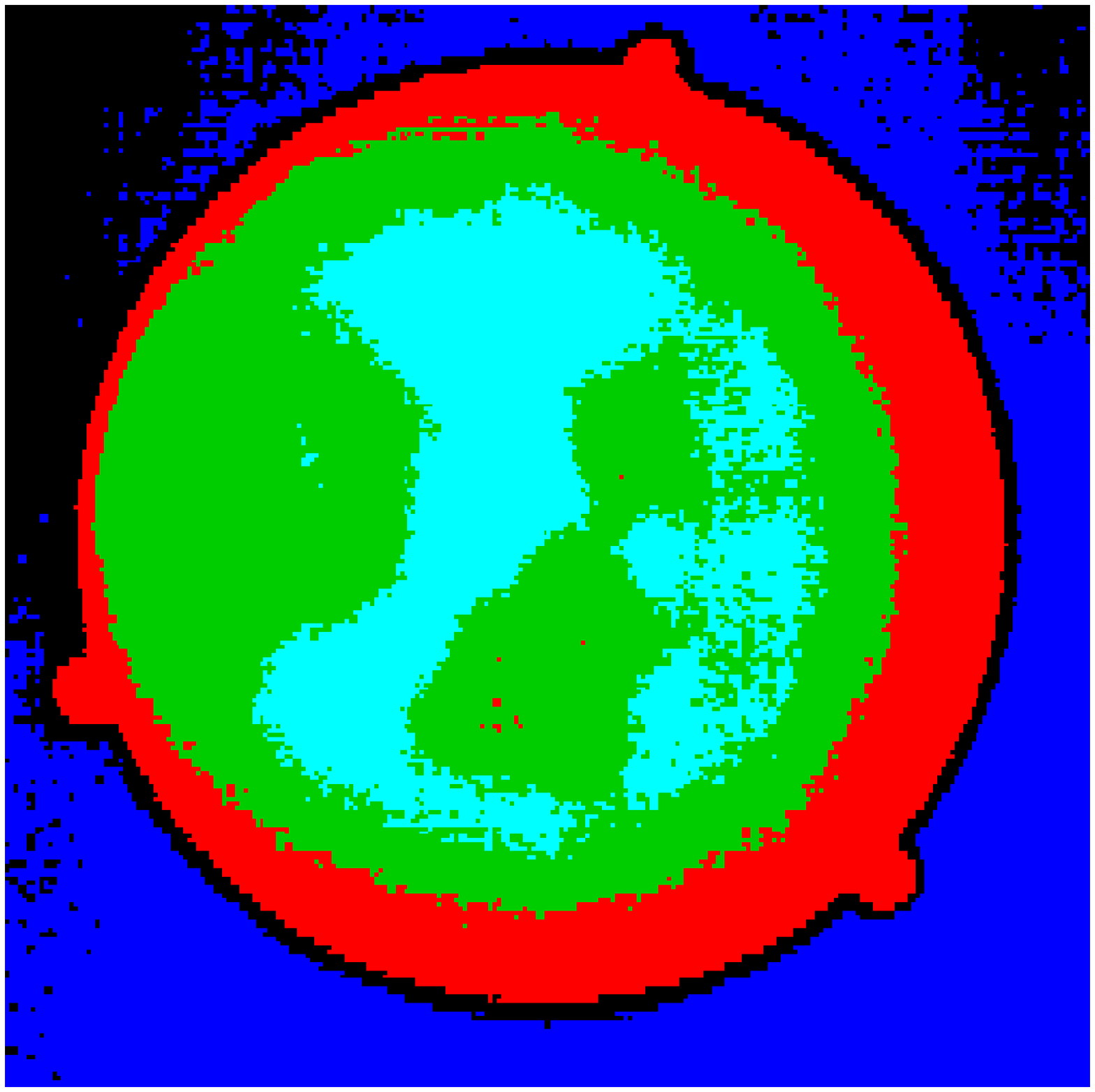}&
  \includegraphics[width=0.3\columnwidth]{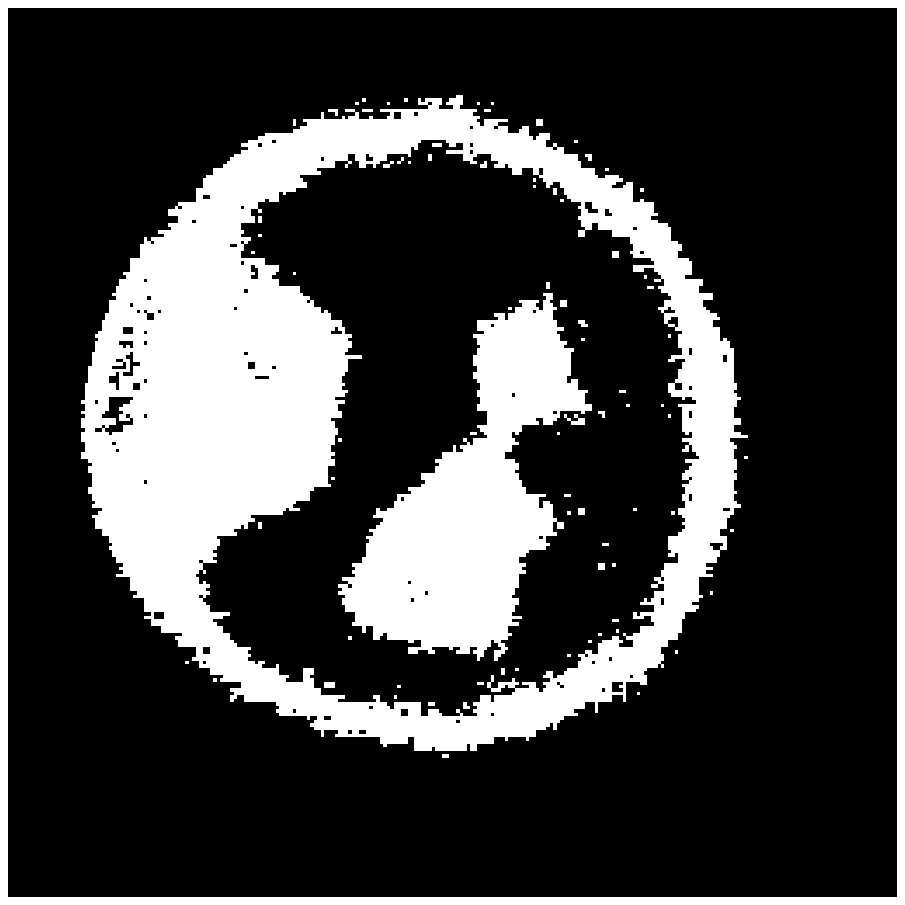}\\
  (a) & (b)\\
  \includegraphics[width=0.3\columnwidth]{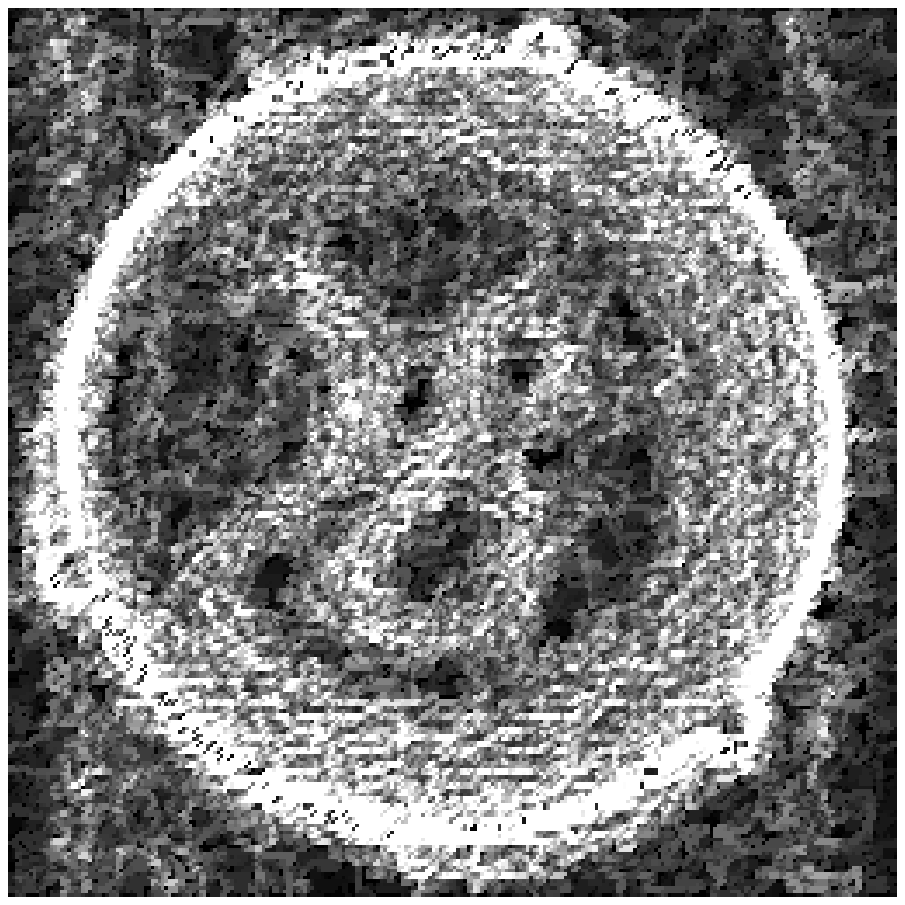}&
  \includegraphics[width=0.3\columnwidth]{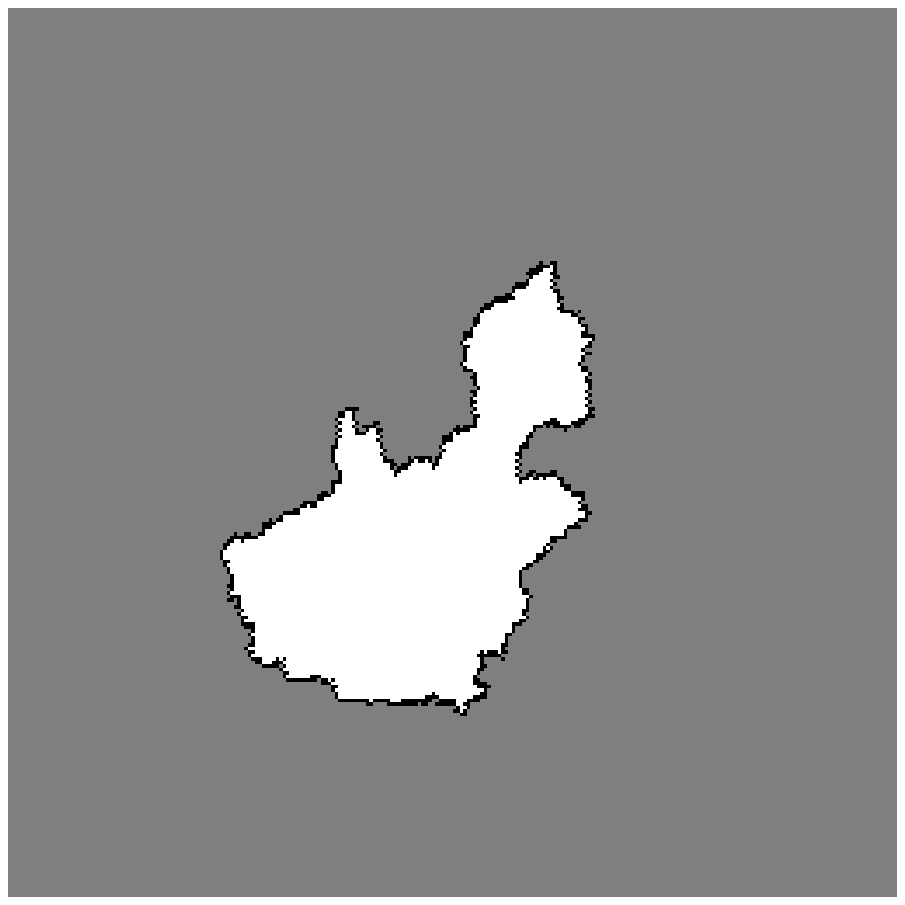}\\
  (c) & (d)\\
  \end{tabular}
\end{center}
\caption{Processes are computed on the filtered image
$\mathbf{\widehat{f}}_{\mathbf{\lambda}}(x)$. (a) Clara
segmentation, (b) Green cluster corresponding to glue used as
markers, (c) Chi-squared gradient, (d) Watershed segmentation.  For
visualization, gradients are scaled by a factor.}
\label{Fig_segmentation_im}
\end{figure}

\begin{figure}
\begin{center}
\begin{tabular}{cc}
  \includegraphics[width=0.3\columnwidth]{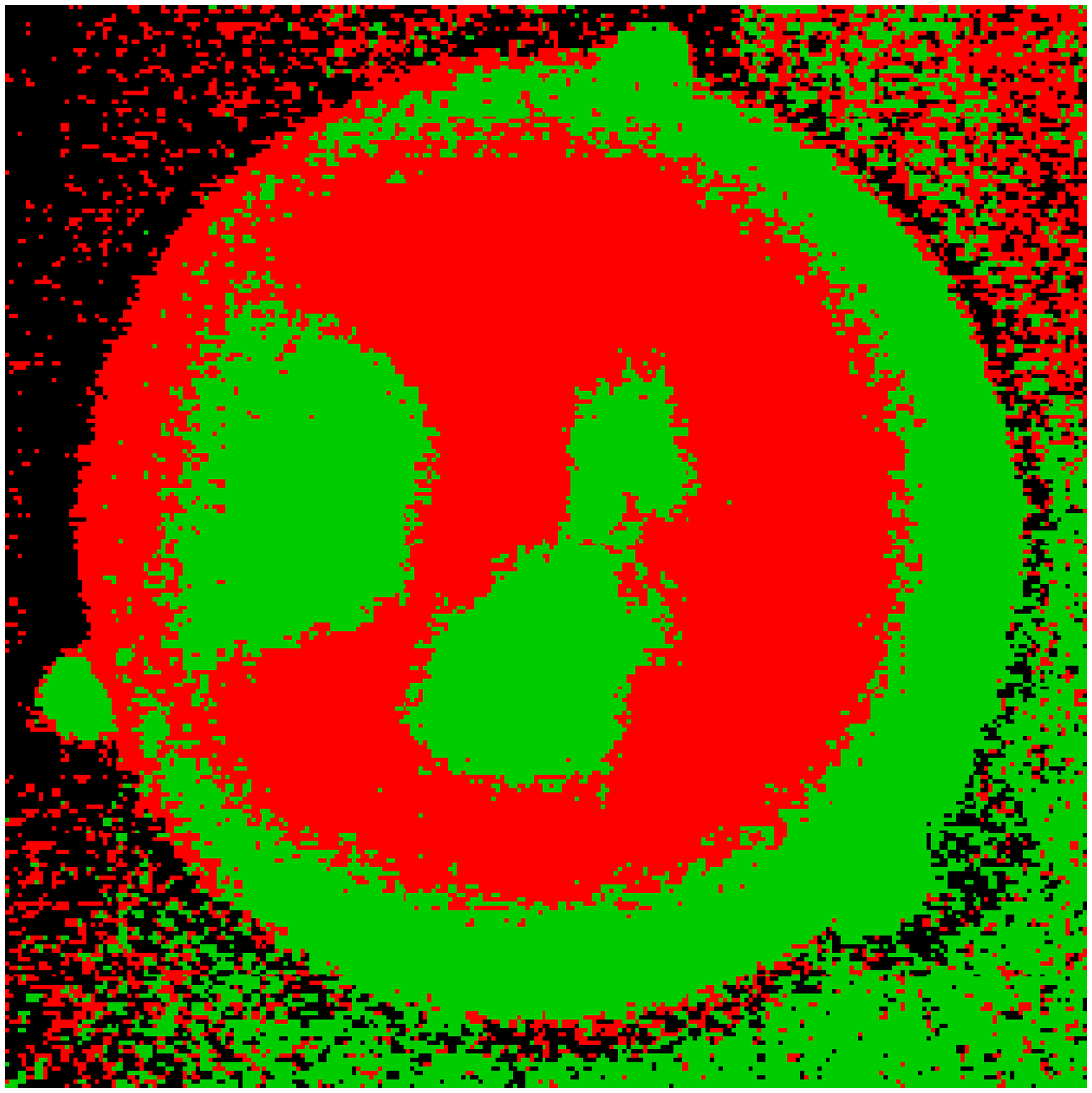}&
  \includegraphics[width=0.3\columnwidth]{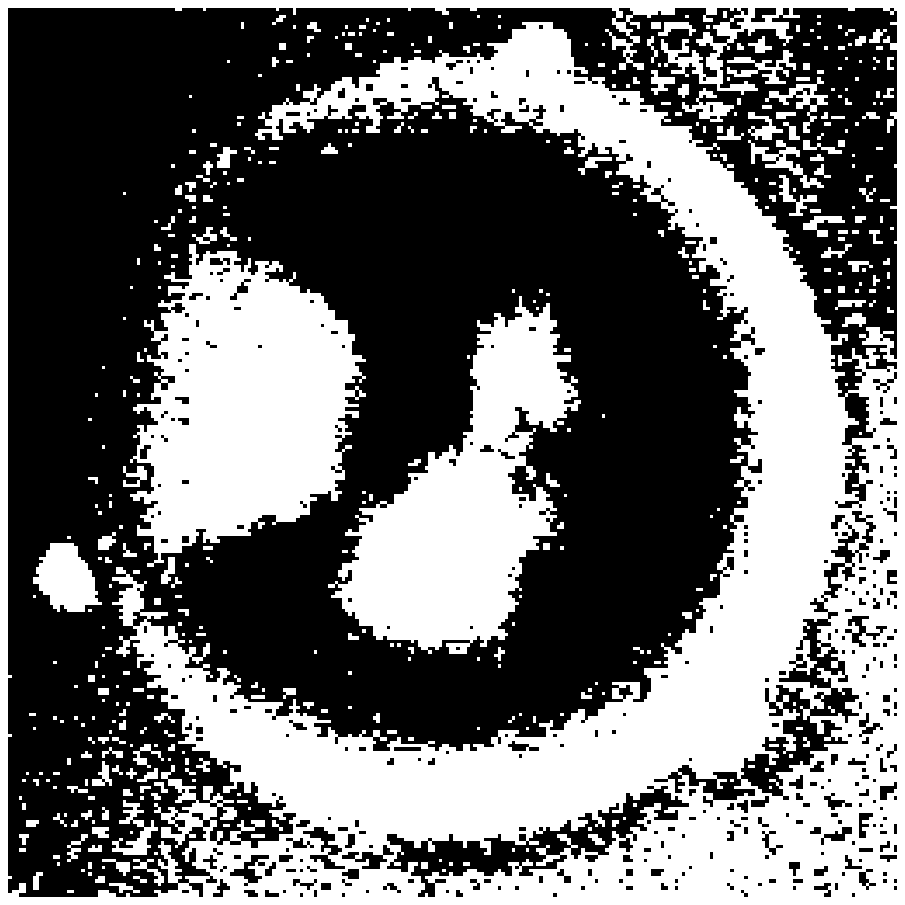}\\
  (a) & (b)\\
  \includegraphics[width=0.3\columnwidth]{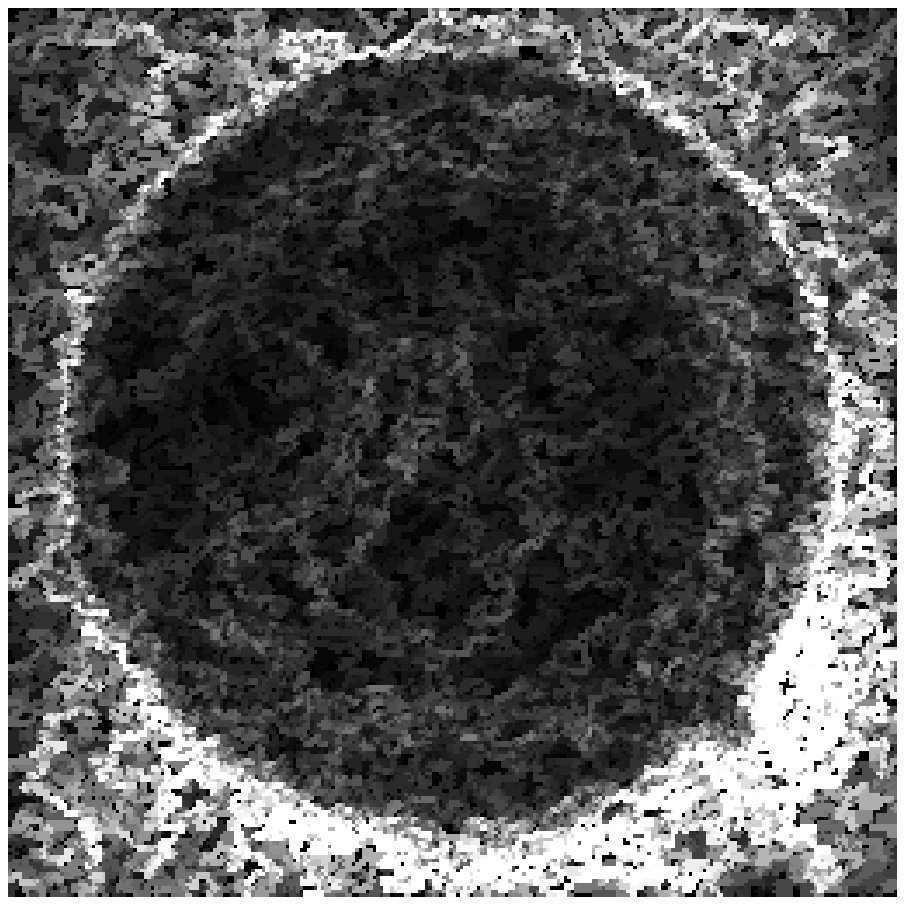}&
  \includegraphics[width=0.3\columnwidth]{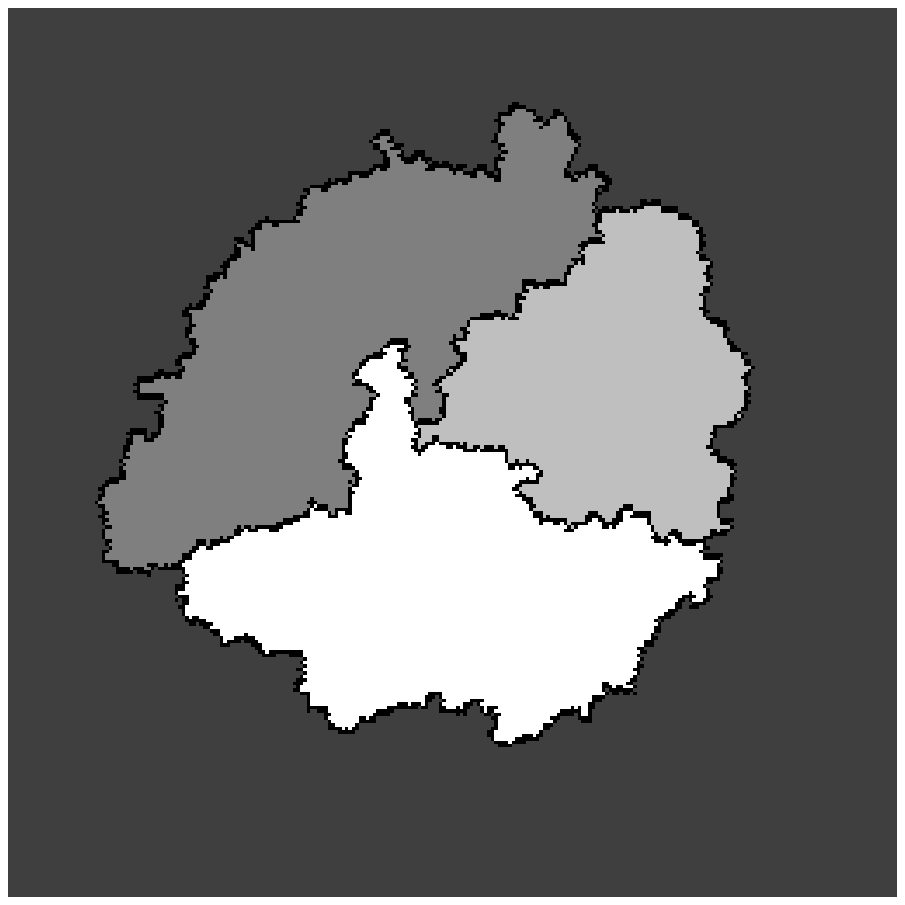}\\
  (c) & (d)\\
  \end{tabular}
\end{center}
\caption{Processes are computed on the factor pixels
$\mathbf{c}^{\mathbf{f}}_{\alpha}(x)$. (a) Clara segmentation, (b)
Green cluster corresponding to glue used as markers, (c) Euclidean
gradient, (d) Watershed segmentation.  For visualization, gradients
are scaled by a factor.} \label{Fig_segmentation_axes_im}
\end{figure}

\section{Data reduction using model approach}

Another way to reduce data is to fit a model on the spectrum of each
pixels vector. The parameters of the model fitted in each pixel are
seen as parametrical cartographies or maps. The whole maps generate
an hyperspectral image $\mathbf{p}(x)=(p_{1}(x),\ldots,p_{M}(x))$.
This approach is advantageous in the way it takes the order of
channels into account. In fact, in FCA channels the order is without
importance.

Moreover, with the hyperspectral image of parameters, a segmentation
can be computed. It is also possible to make parameters orthogonal
with a Principal Component Analysis and then generating a
segmentation.

Principal Component Analysis (PCA) is used here instead of FCA,
because it is possible to compute it on hyperspectral images with
negative values, which can be the case for some parameters. PCA
gives factorial axes with factors building another hyperspectral
image:
$\mathbf{c}^{\mathbf{p}}_{\beta}(x)=(c^{\mathbf{p}}_{\beta_{1}}(x),\ldots,c^{\mathbf{p}}_{\beta_{M}}(x))$.

For this thermographic sequence, a linear model is fitted on the
image after removing the 10 first channels which correspond to a
transitory phenomenon (fig. \ref{Fig_spectres}). The linear model
$y=ax+b$ has two parameters, the slope $a$ and the intercept $b$. On
the first 10 channels we have defined another parameter, called rise
$m$, as the maximum amplitude on these 10 channels:
\begin{equation}\label{eq_montee}
     m(x) = \max_{\substack{j\in[1:10]}}(\widehat{f}_{\lambda_{j}}(x))-
    \min_{\substack{j\in[1:10]}}(\widehat{f}_{\lambda_{j}}(x))).
\end{equation}

\begin{figure}
\begin{center}
\begin{tabular}{cc}
  \includegraphics[width=0.3\columnwidth]{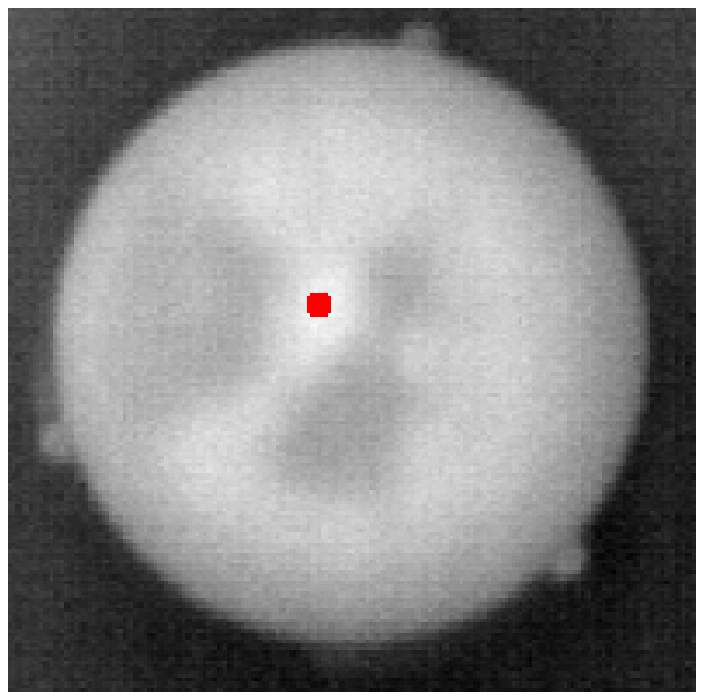}&
  \includegraphics[width=0.33\columnwidth]{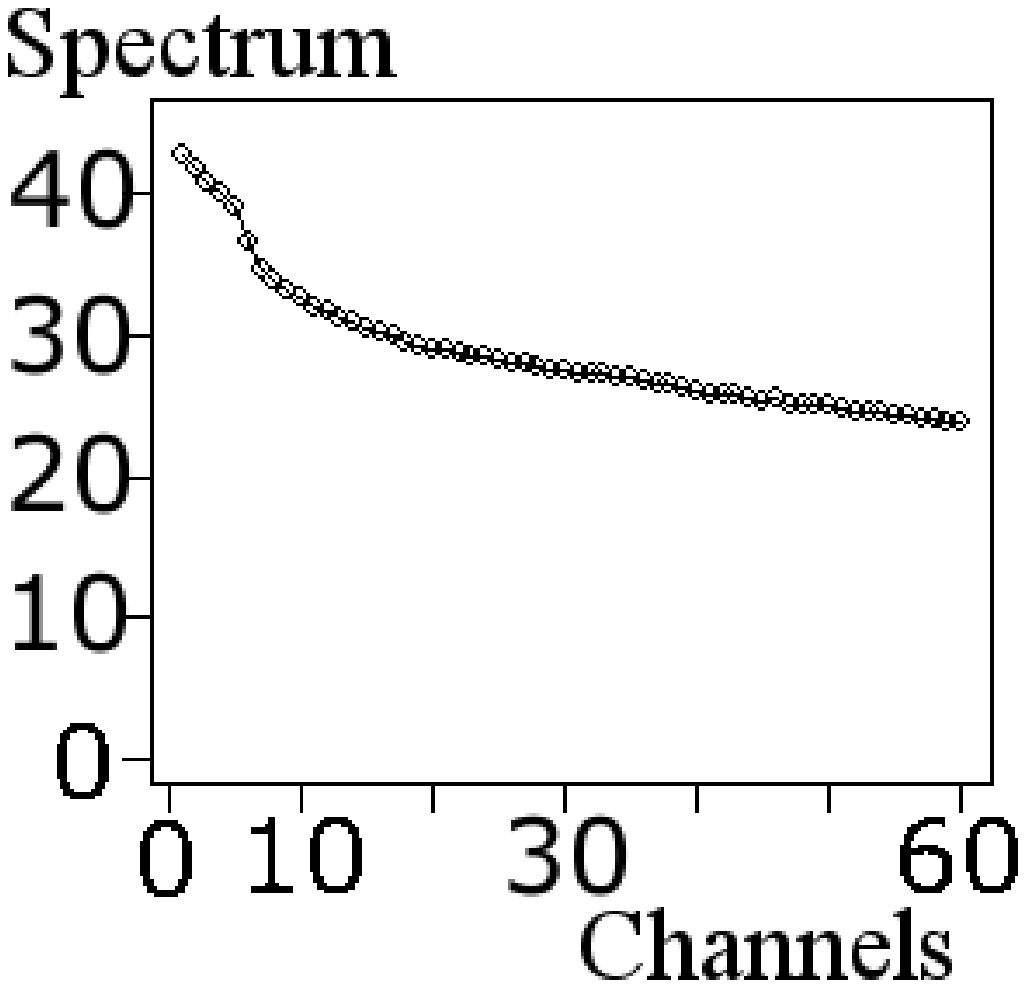}\\
  (a) & (b) \\
  \includegraphics[width=0.3\columnwidth]{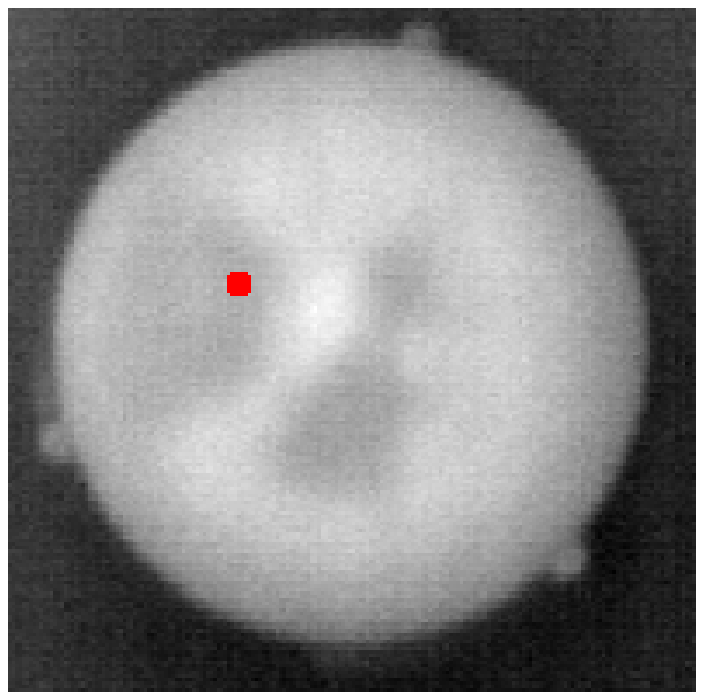}&
  \includegraphics[width=0.33\columnwidth]{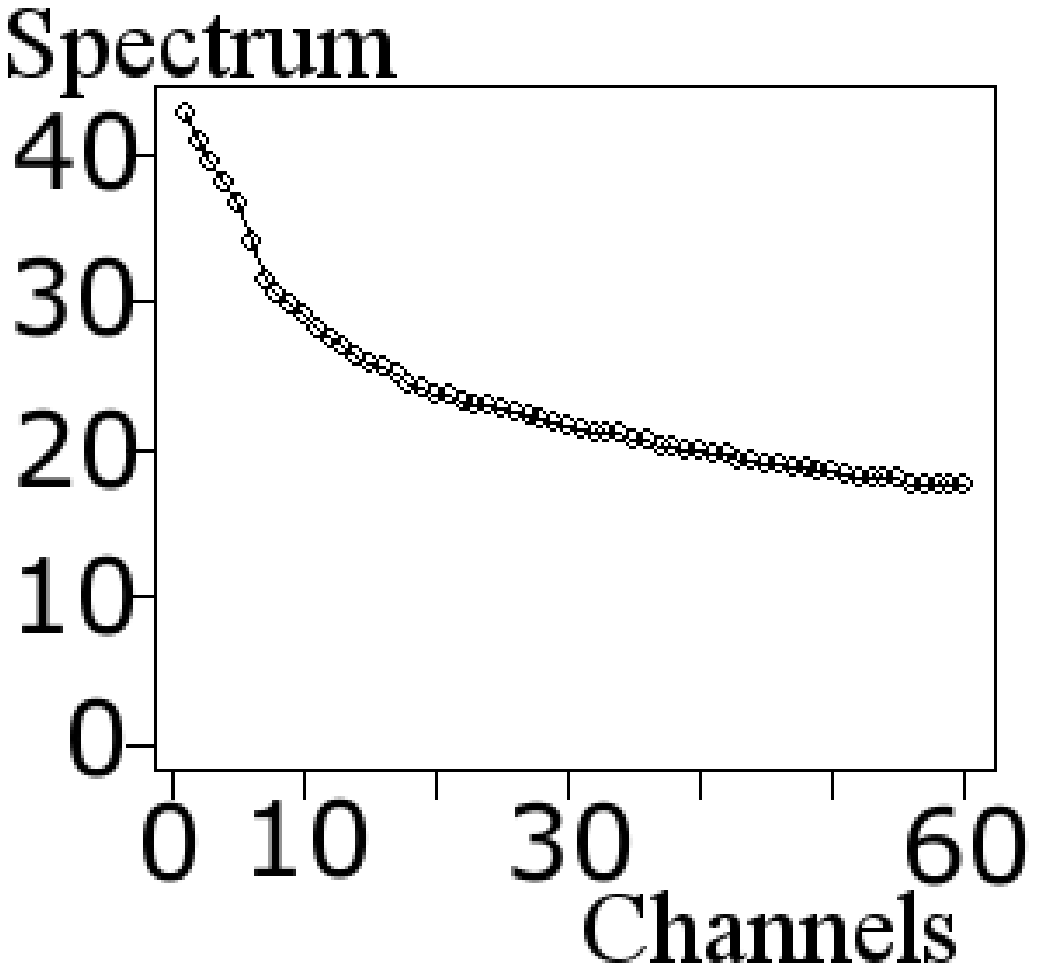}\\
 (c) & (d)\\
  \end{tabular}
\end{center}
  \caption{(a) Point outside glue where is measured spectrum (b),
  (c) point inside glue and corresponding spectrum (d).}
  \label{Fig_spectres}
\end{figure}

Then with the linear model we obtain images as maps of the
parameters, which can be orthogonalised by PCA (fig
\ref{Fig_para_axes_ACP_para}). In this case the different axes
represent the following inertia ratios $97.24\%$, $2.10\%$ and
$0.66\%$.
\begin{figure}
\begin{center}
\begin{tabular}{ccc}
  \includegraphics[width=0.3\columnwidth]{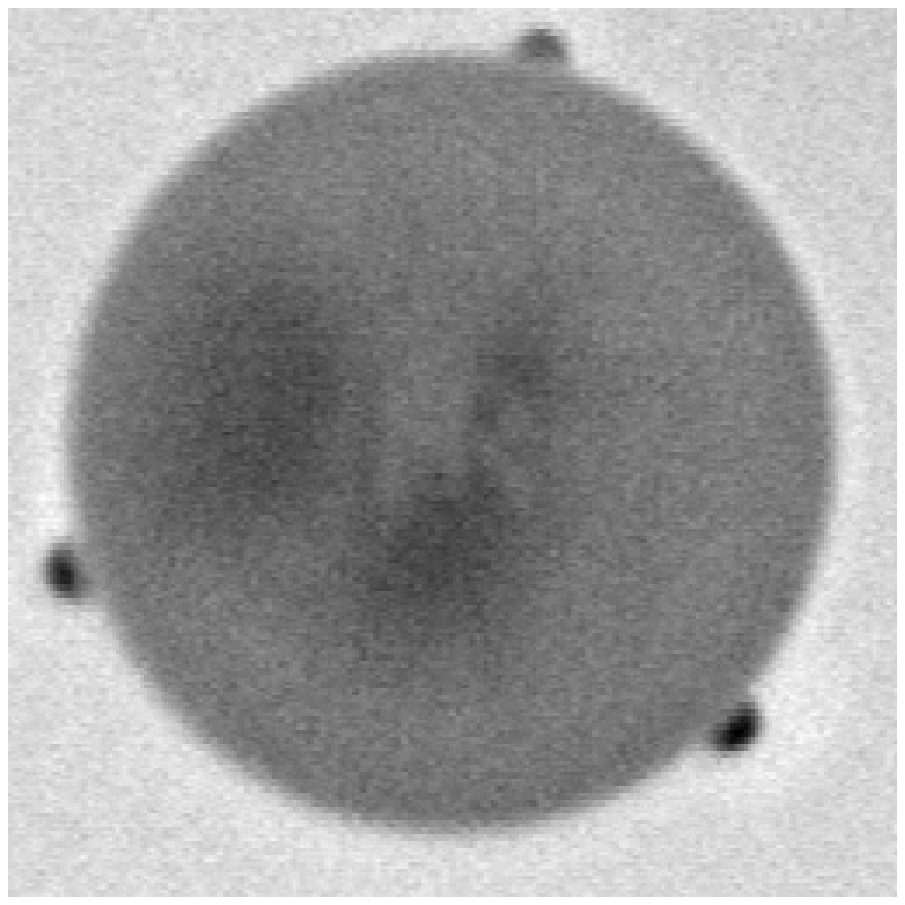}&
  \includegraphics[width=0.3\columnwidth]{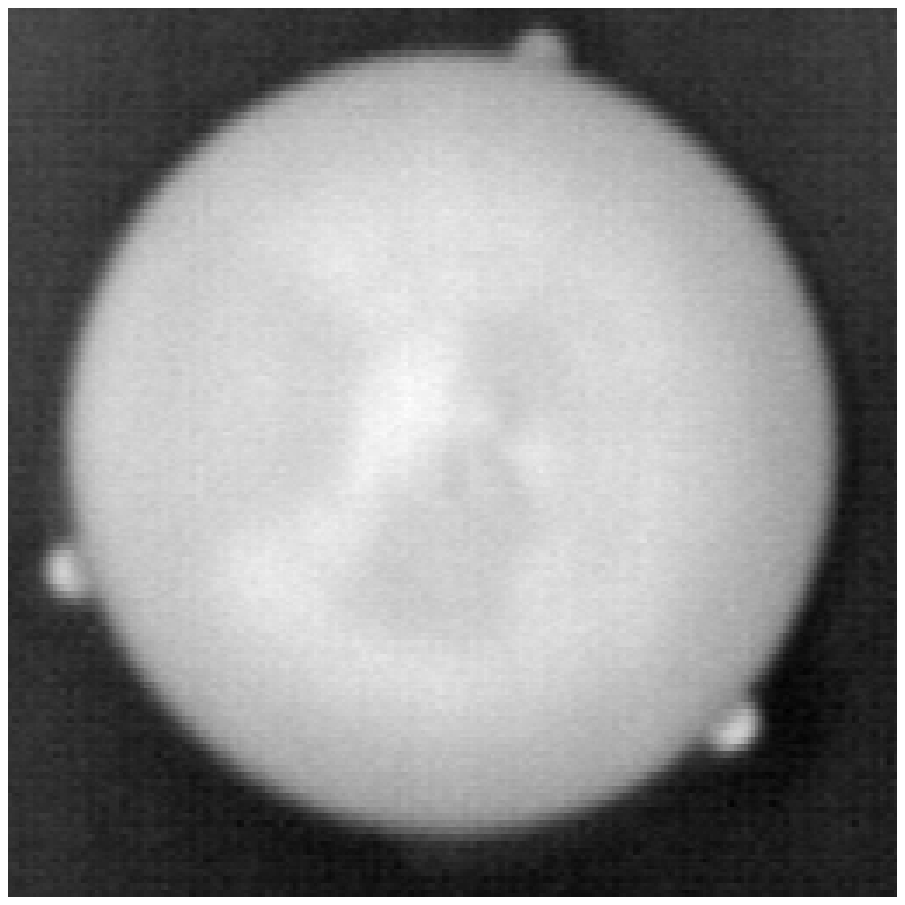}&
  \includegraphics[width=0.3\columnwidth]{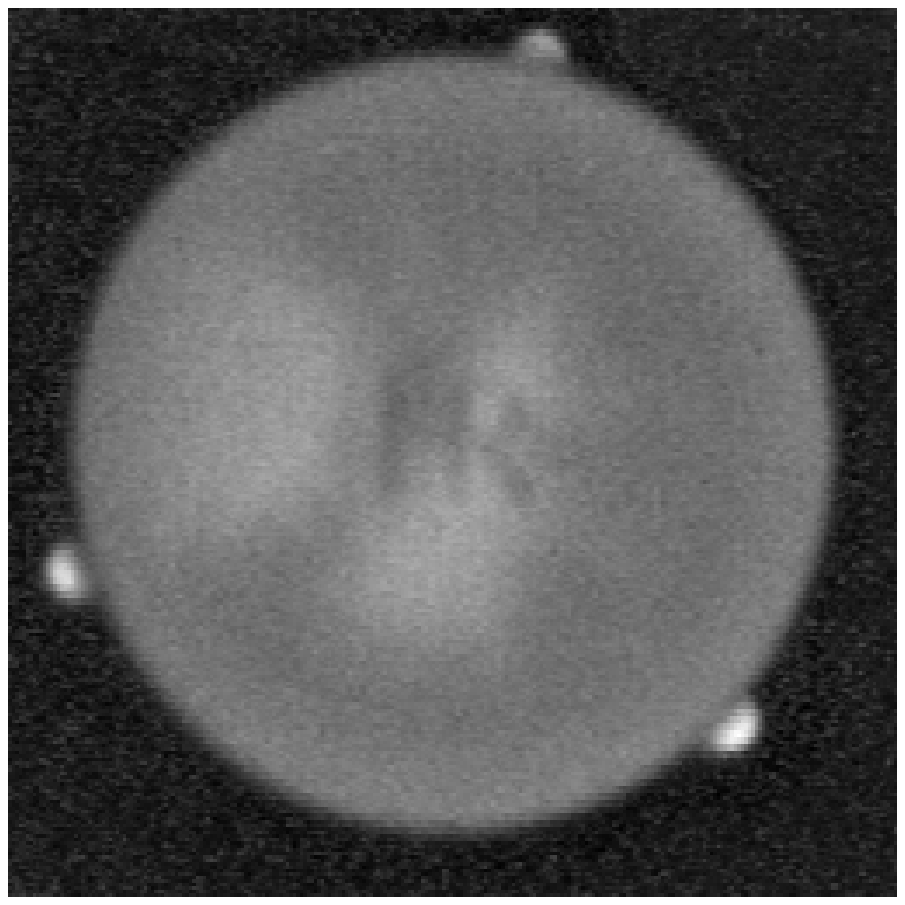}\\
  $p_{1} = a$ & $p_{2} = b$ & $p_{3} = m$\\
  \includegraphics[width=0.3\columnwidth]{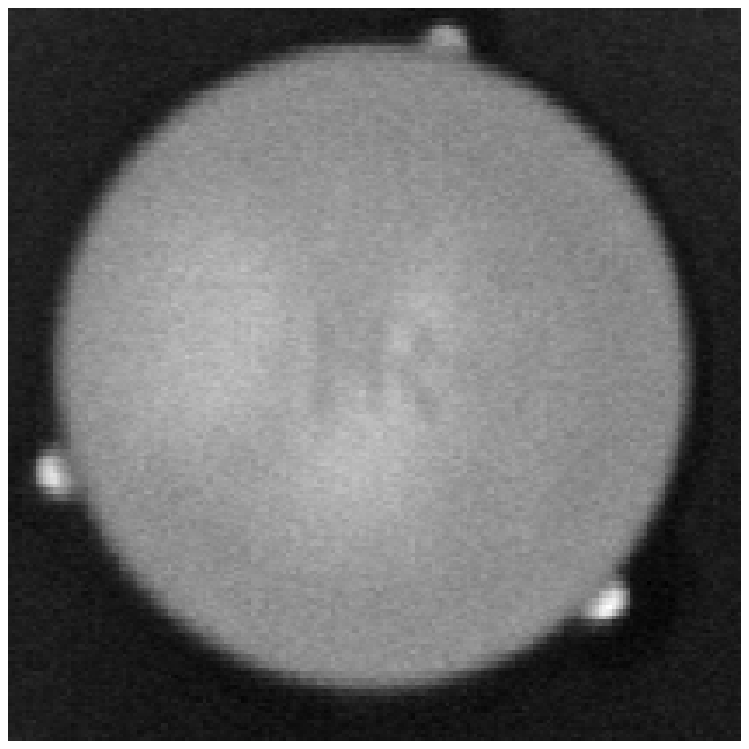}&
  \includegraphics[width=0.3\columnwidth]{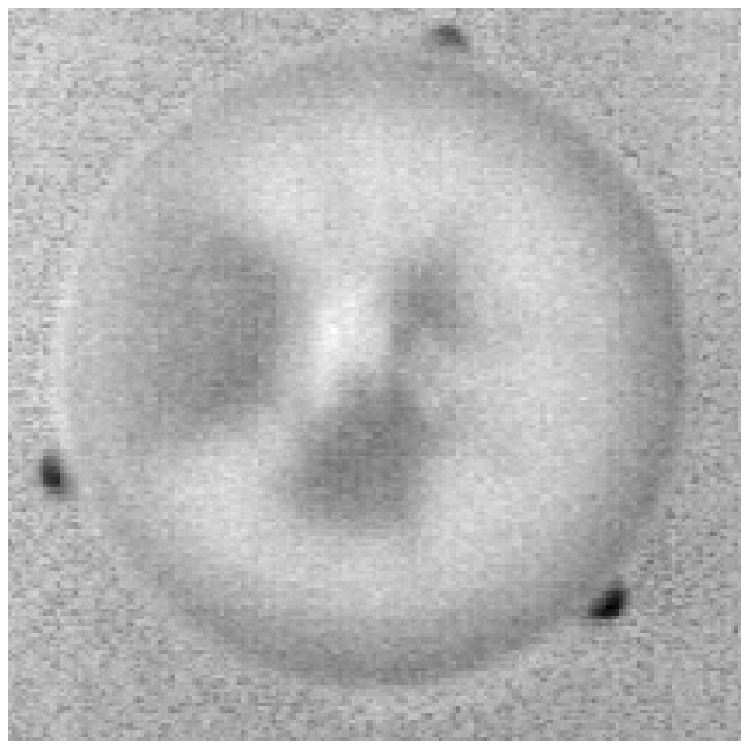}&
  \includegraphics[width=0.3\columnwidth]{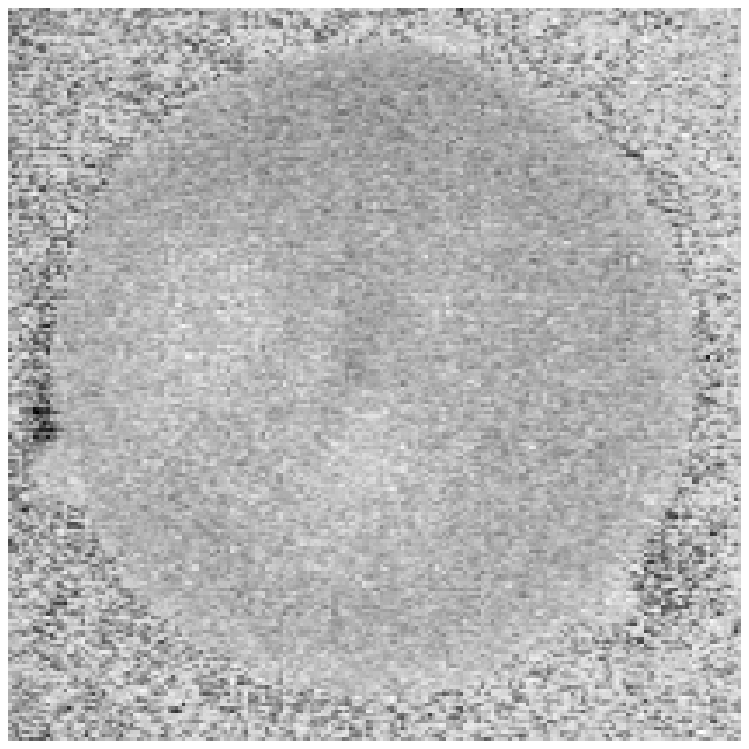}\\
  axis 1 $c^{\mathbf{p}}_{\beta_{1}}$ & axis 2 $c^{\mathbf{p}}_{\beta_{2}}$ & axis 3 $c^{\mathbf{p}}_{\beta_{3}}$\\
  \end{tabular}
\end{center}
  \caption{\emph{Maps of the parameters of the linear model, PCA factors of the parameters on axis 1, 2 and 3..}}
  \label{Fig_para_axes_ACP_para}
\end{figure}

\subsection{Segmentation of $\mathbf{p}(x)$ or $\mathbf{c}^{\mathbf{p}}_{\beta}(x)$}

Different approaches of segmentation are tested on parameters
$\mathbf{p}(x)$ or on parameters factors
$\mathbf{c}^{p}_{\beta}(x)$. Markers are computed again by the
cluster method "Clara" on the parameters and the image of factors .
First, a clustering is processed on both images. The cluster
corresponding to the lid center is selected because the glue is on
the lid center. A second clustering "Clara" is made on the selected
cluster. Non selected pixels are aggregated to the cluster of the
largest size, because cluster corresponding to glue has the smallest
size (fig. \ref{Fig_marqueurs_clara_para_ACP_para}).

\begin{figure}
\begin{center}
\begin{tabular}{cc}
  \includegraphics[width=0.3\columnwidth]{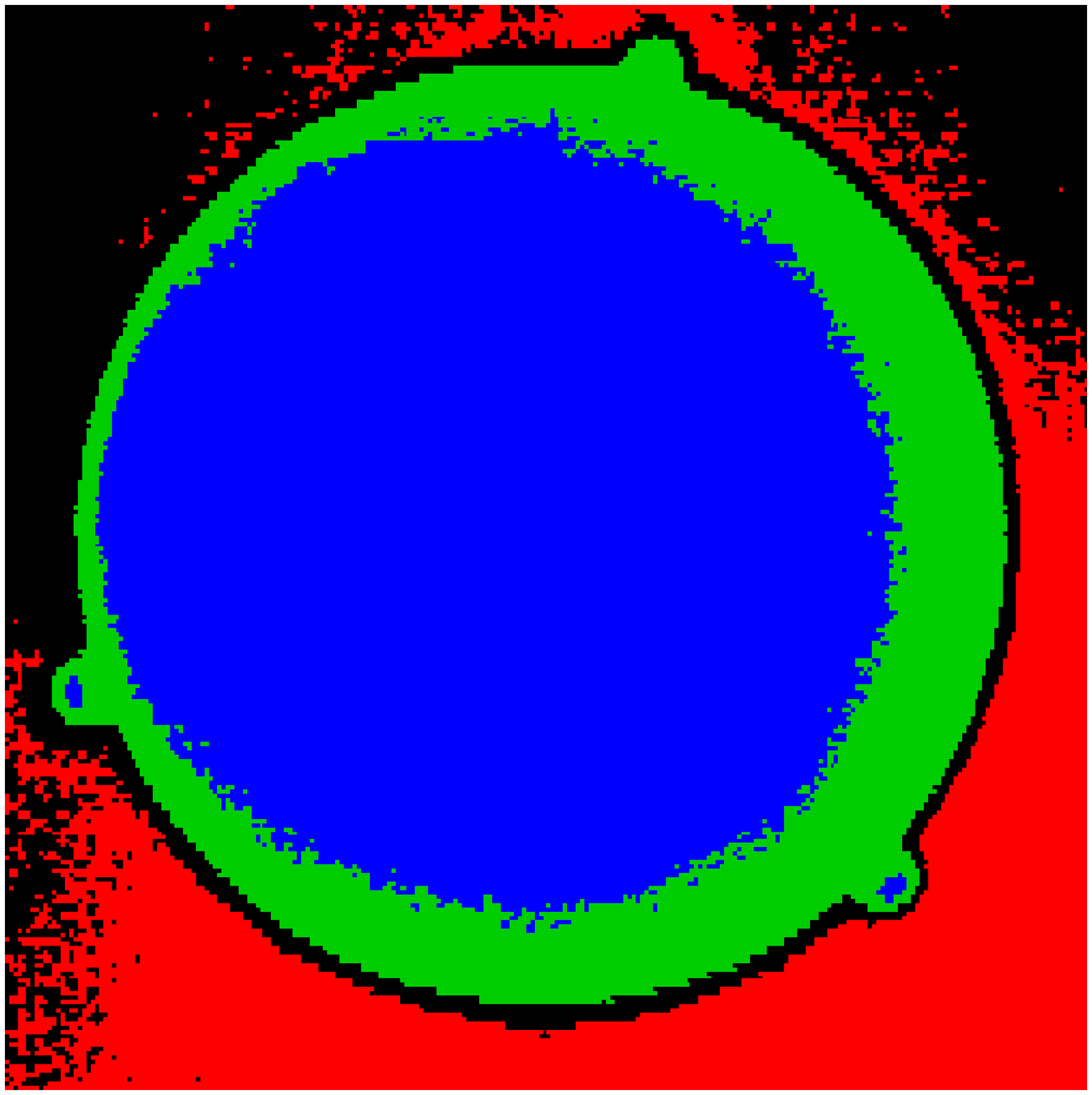}&
  \includegraphics[width=0.3\columnwidth]{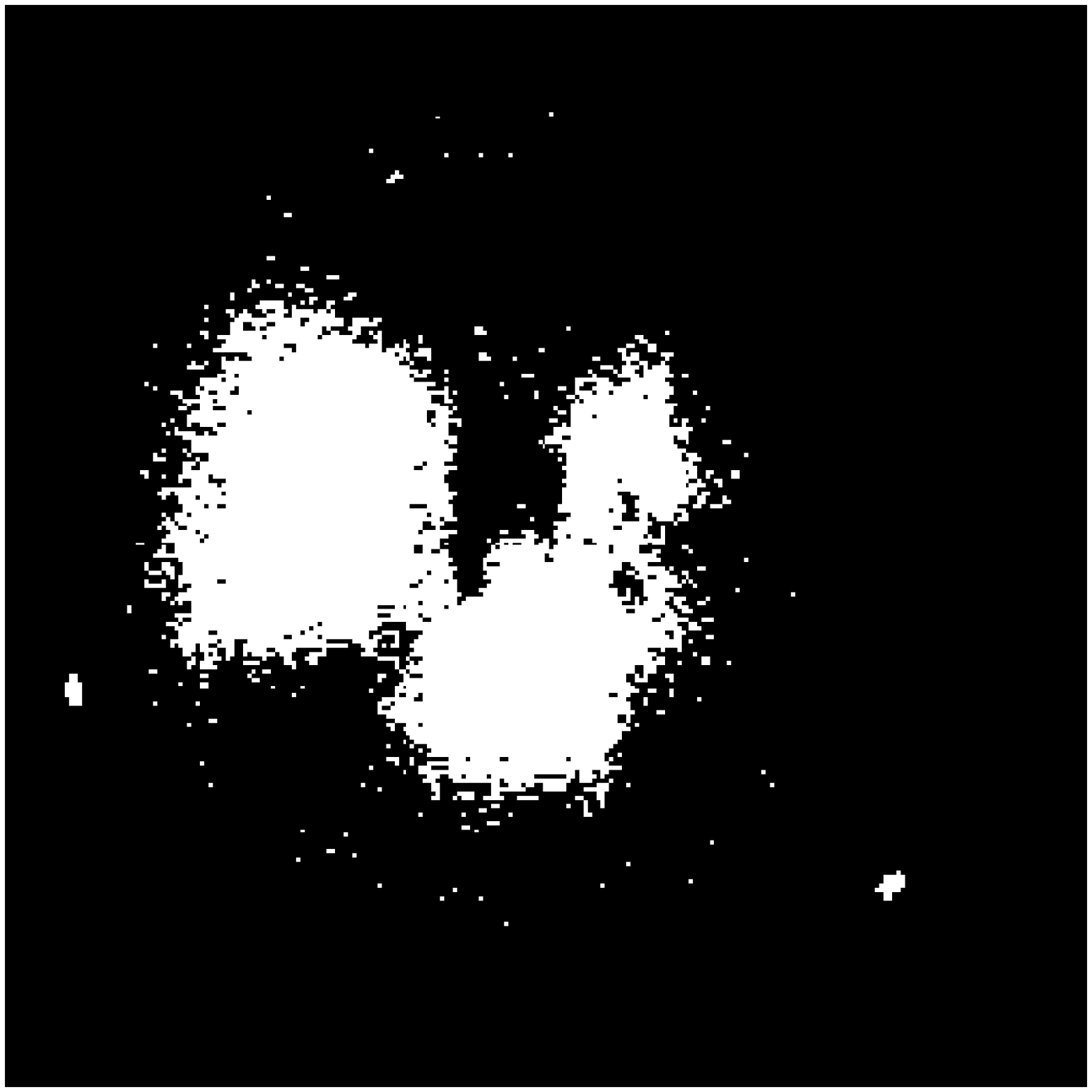}\\
  (a) & (b)\\
  \includegraphics[width=0.3\columnwidth]{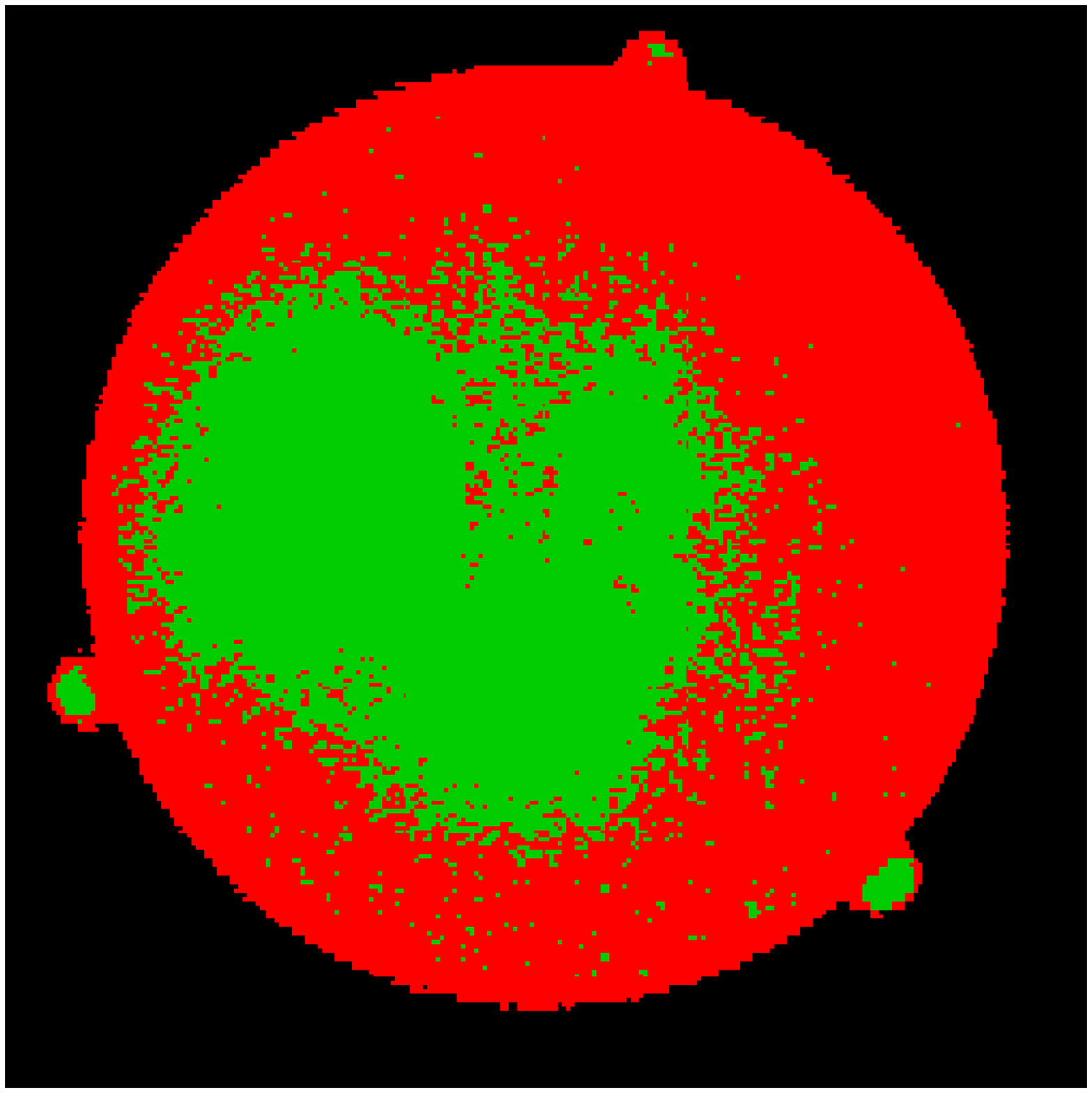}&
  \includegraphics[width=0.3\columnwidth]{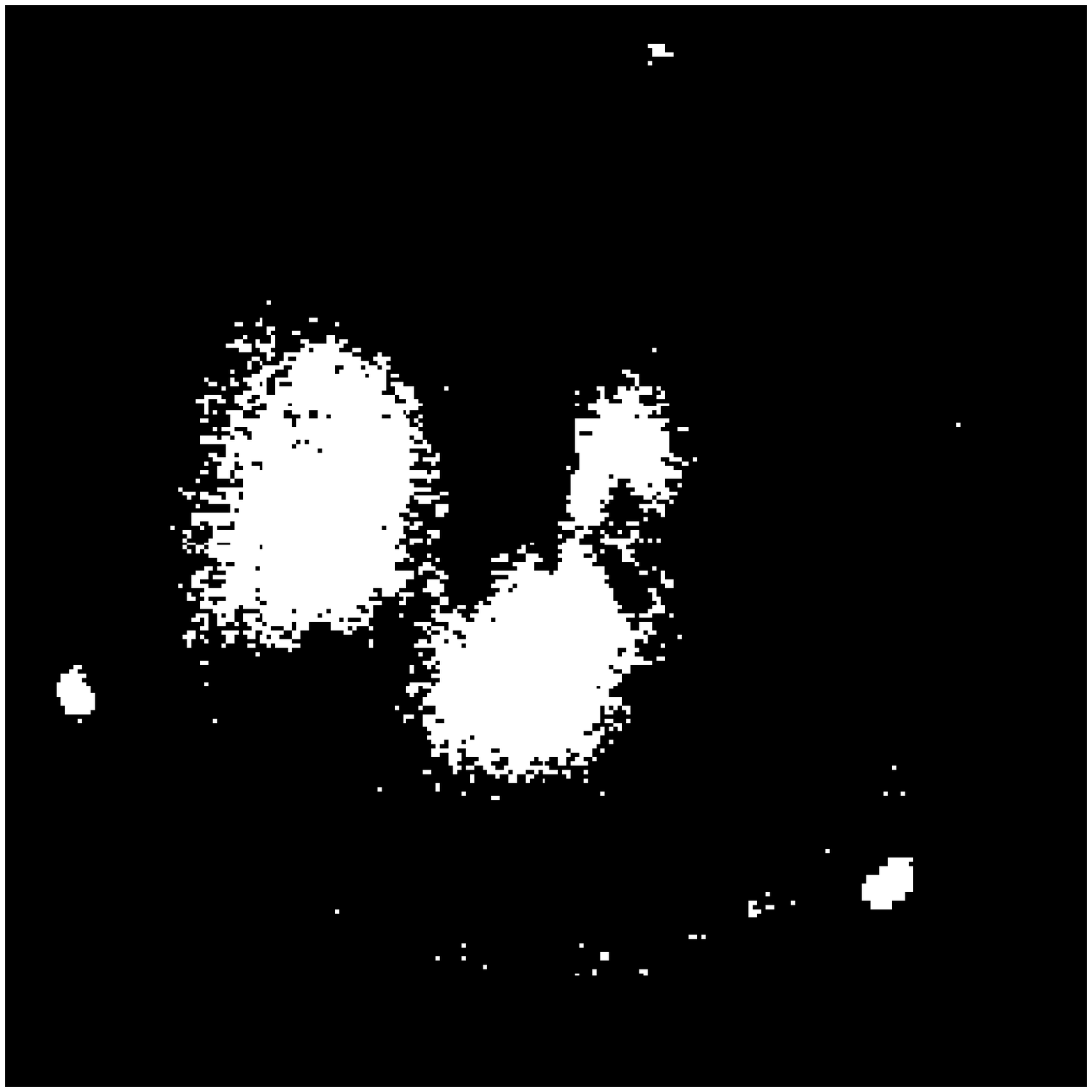}\\
  (c) & (d)\\
  \end{tabular}
\end{center}
\caption{(a) Clara segmentation in 4 clusters on parameter $b$, blue
cluster is selected, (b) Clara on parameters $a$ and $m$:
parameters markers, (c) Clara segmentation in 3 clusters on first
axis $c^{\mathbf{p}}_{\beta_{1}}$, green cluster is selected, (d)
Clara on axes 2 and 3 ($c^{\mathbf{p}}_{\beta_{2}}$,
$c^{\mathbf{p}}_{\beta_{3}}$): parameters factors
markers.}\label{Fig_marqueurs_clara_para_ACP_para}
\end{figure}

To improve the quality of gradients, each image channel is again
leveled using as image markers the corresponding channels, filtered
by a gaussian kernel. Then, several gradients are tested. The
morphological gradient on each channel $g$, the supremum and the sum
of morphological gradients, $\nabla_{\vee}$ and $\nabla_{+}$, are
evaluated on the image of parameters $\mathbf{p}(x)$ and on the
image of PCA factors parameters
$\mathbf{c}^{\mathbf{p}}_{\beta}(x)$. Besides the Euclidean distance
in PCA factorial space is equivalent to the Mahalanobis distance in
parameters space. Therefore, the Mahalanobis distance based gradient
$\nabla_{M}\mathbf{p}(x)$ is performed on the parameters
$\mathbf{p}(x)$ to ensure the equal statistical value for the
different parameters. For the PCA factors parameters
$\mathbf{c}^{\mathbf{p}}_{\beta}(x)$, the most indicated is the
Euclidean distance based gradient.

\subsubsection{Watershed on parameters}

The morphological gradient is computed on each channel of image
$\mathbf{p}(x)$. Markers (fig.
\ref{Fig_marqueurs_clara_para_ACP_para} (b)) are filtered by opening
with an hexagonal structuring element of size 5. With the gradient
and the markers, a watershed segmentation is obtained for each
parameter.

\begin{figure}
\begin{center}
\begin{tabular}{ccc}
  \includegraphics[width=0.3\columnwidth]{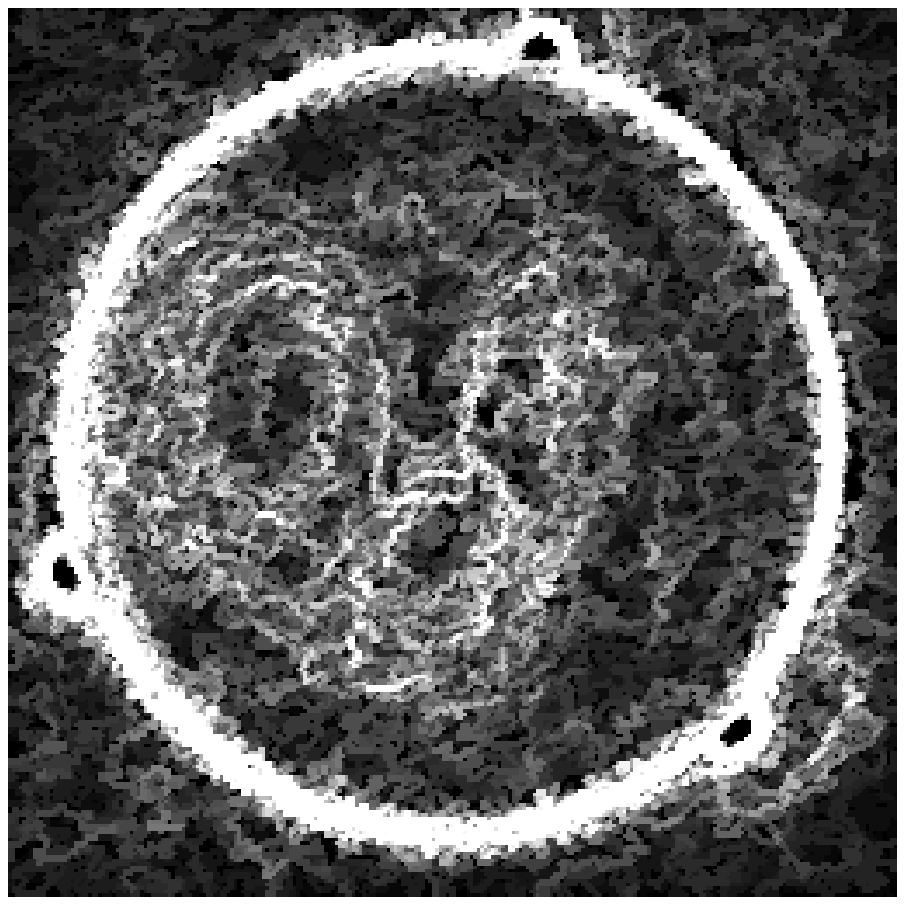}&
  \includegraphics[width=0.3\columnwidth]{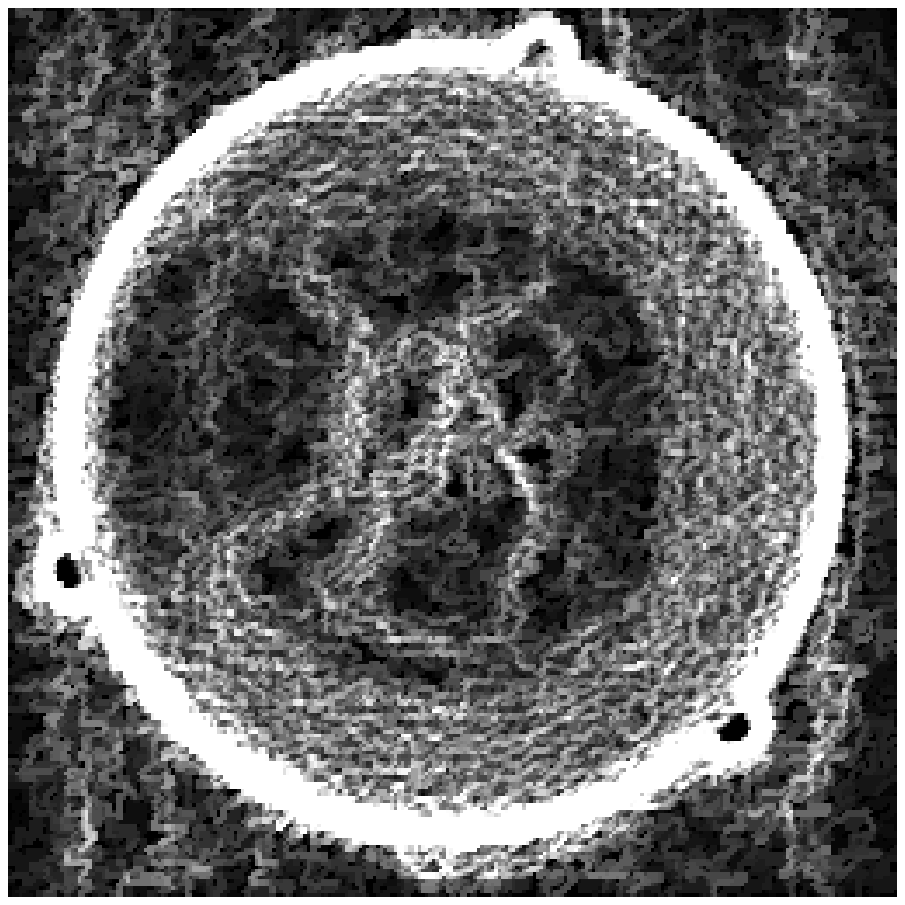}&
  \includegraphics[width=0.3\columnwidth]{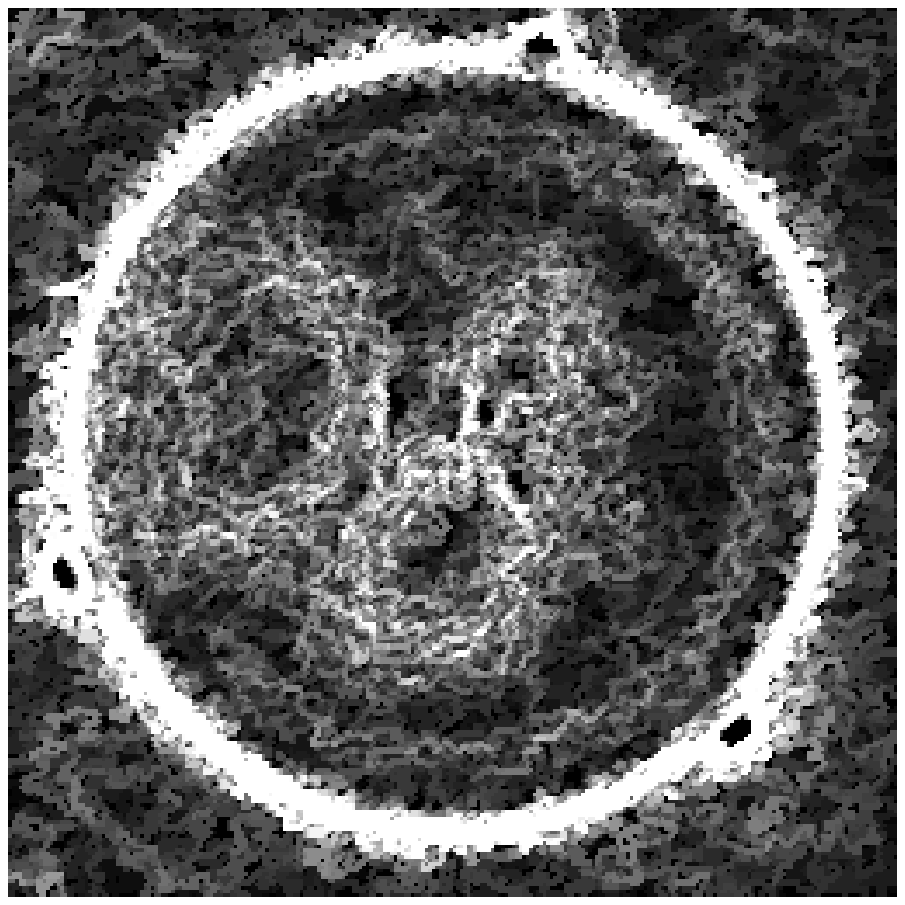}\\
  \includegraphics[width=0.3\columnwidth]{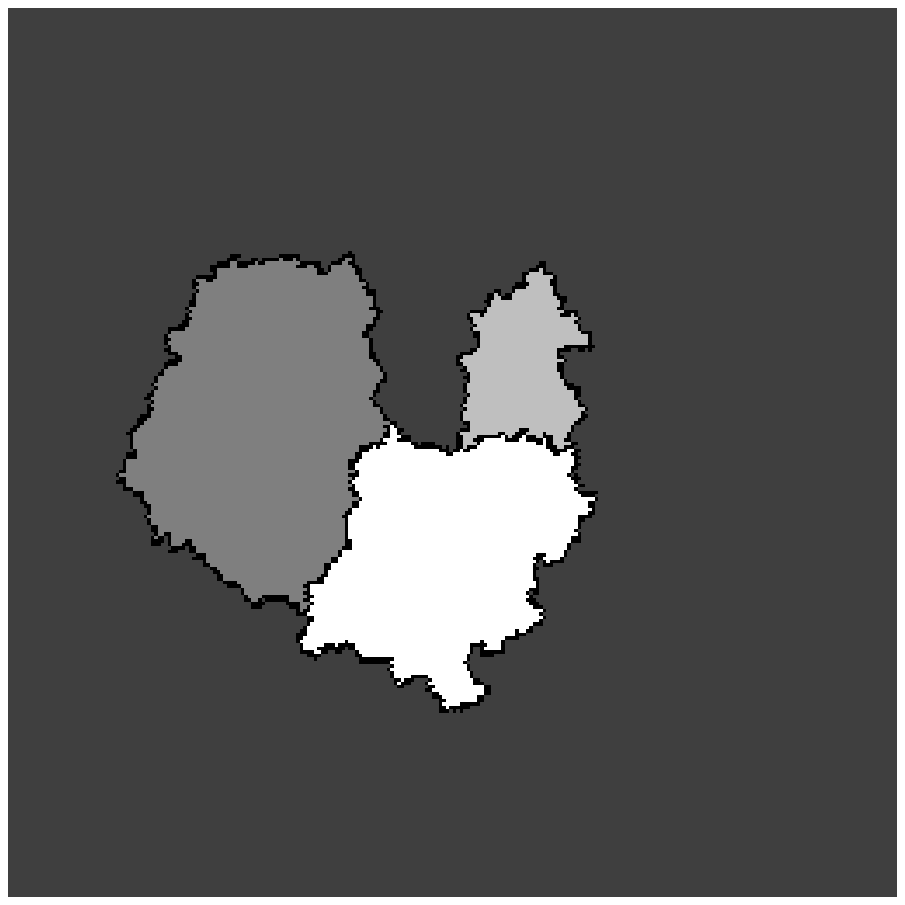}&
  \includegraphics[width=0.3\columnwidth]{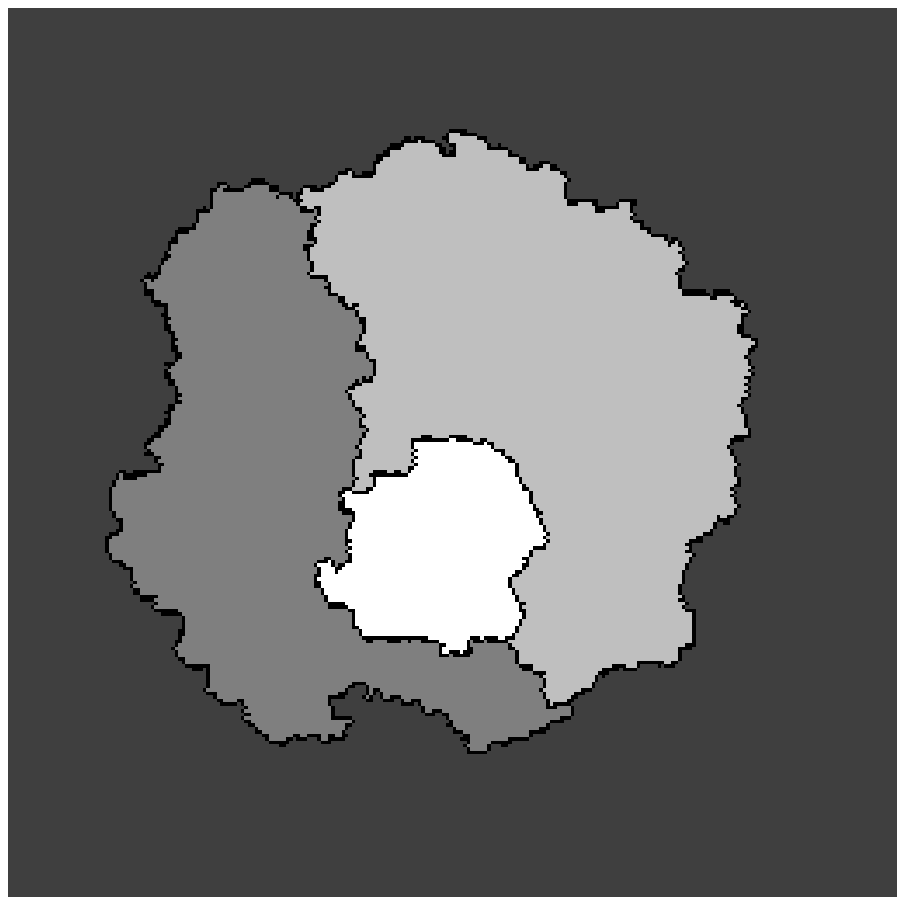}&
  \includegraphics[width=0.3\columnwidth]{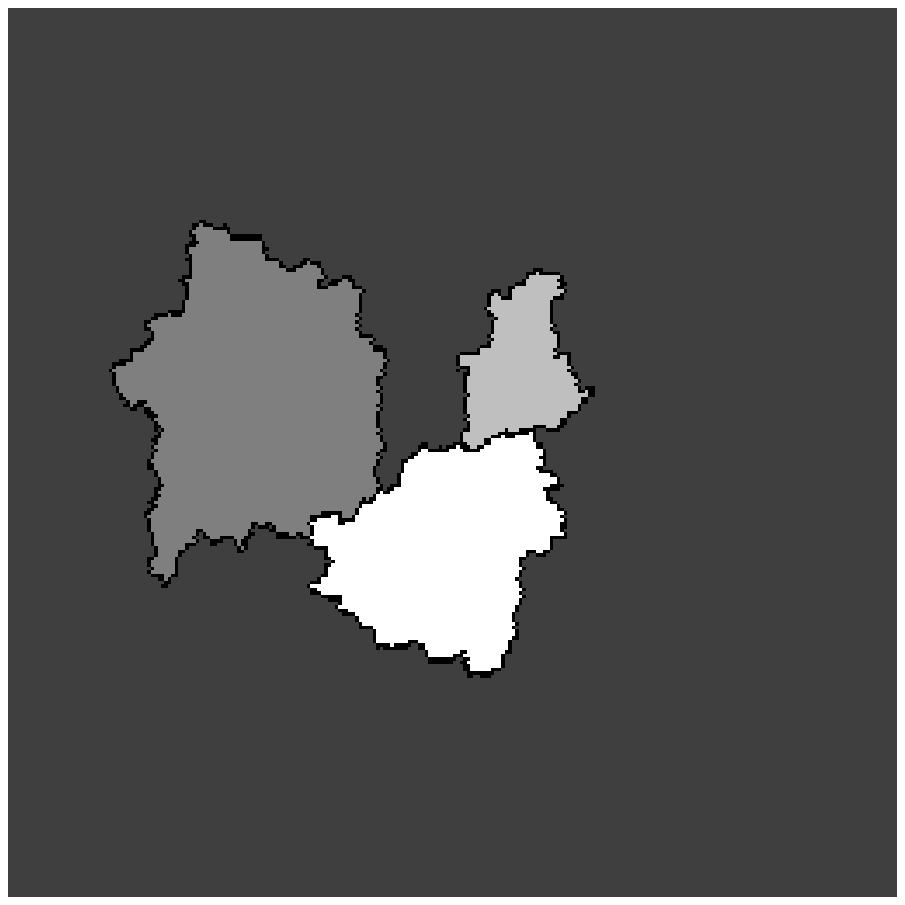}\\
  $a$ & $b$ & $m$\\
\end{tabular}
\end{center}
  \caption{First line: morphological gradient, second line: watershed on parameters: slope $a$, intercept $b$ and rise $m$.
  For visualization, gradients are scaled by a factor. }
  \label{Fig_segmentation_gradient_morpho_para}
\end{figure}

Comparing to the reference, segmentations with morphological
gradients on parameters slope $a$ and rise $m$ are good. However,
the segmentation on intercept $b$ is not satisfactory, due to the
leaks on the gradient. Although these segmentations are pertinent,
they are only marginal segmentations, i.e. only one parameter is
taken into account in one segmentation. Therefore, we have tested
the use of a vectorial gradient.

A metric-based gradient with the Mahalanobis distance
$\nabla_{M}\mathbf{p}(x)$, and a gradient supremum of channels
gradient $\nabla_{\vee}\mathbf{p}(x)$ are tested on the parameters
image (fig. \ref{Fig_segmentation_gradient_euclidien_et_sup_para}).
In this case the gradient sum $\nabla_{+}\mathbf{p}(x)$ is
approximatively equivalent to the supremum gradient because it is
not possible to define specific weights for the parameters. For each
case, the same markers are used. Moreover, to compute the
Mahalanobis distance, we assume that parameters are uncorrelated.
\begin{figure}
\begin{center}
\begin{tabular}{cc}
  \includegraphics[width=0.3\columnwidth]{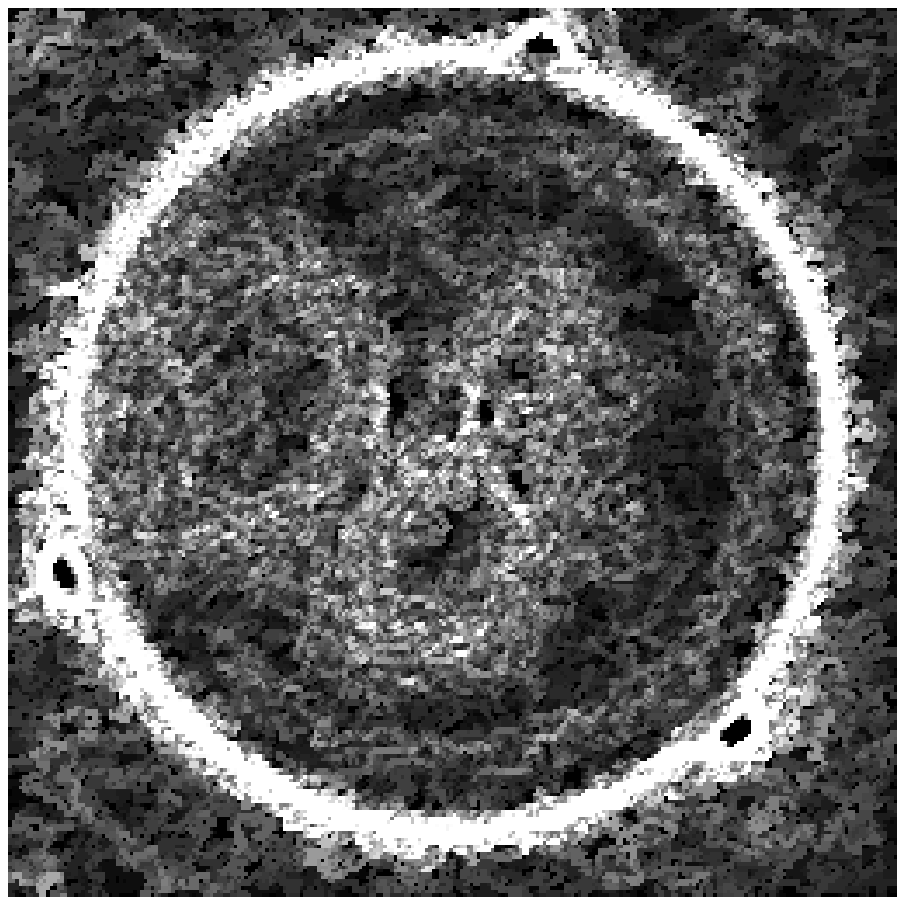}&
  \includegraphics[width=0.3\columnwidth]{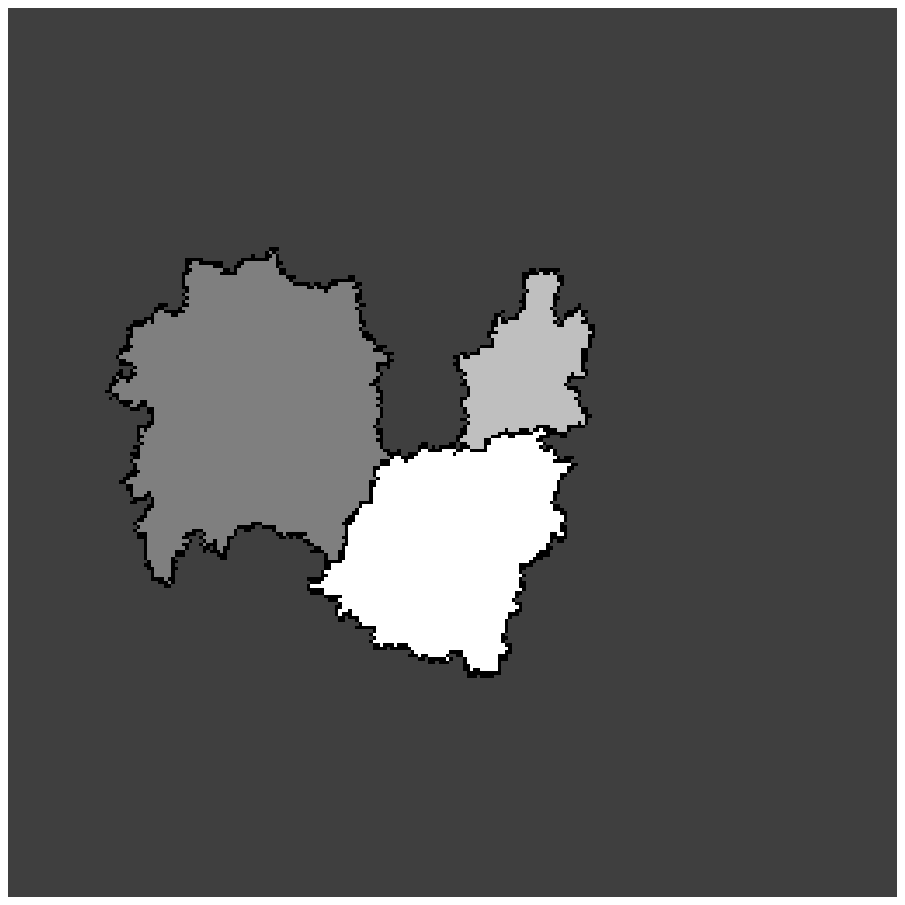}\\
  (a) & (b)\\
  \includegraphics[width=0.3\columnwidth]{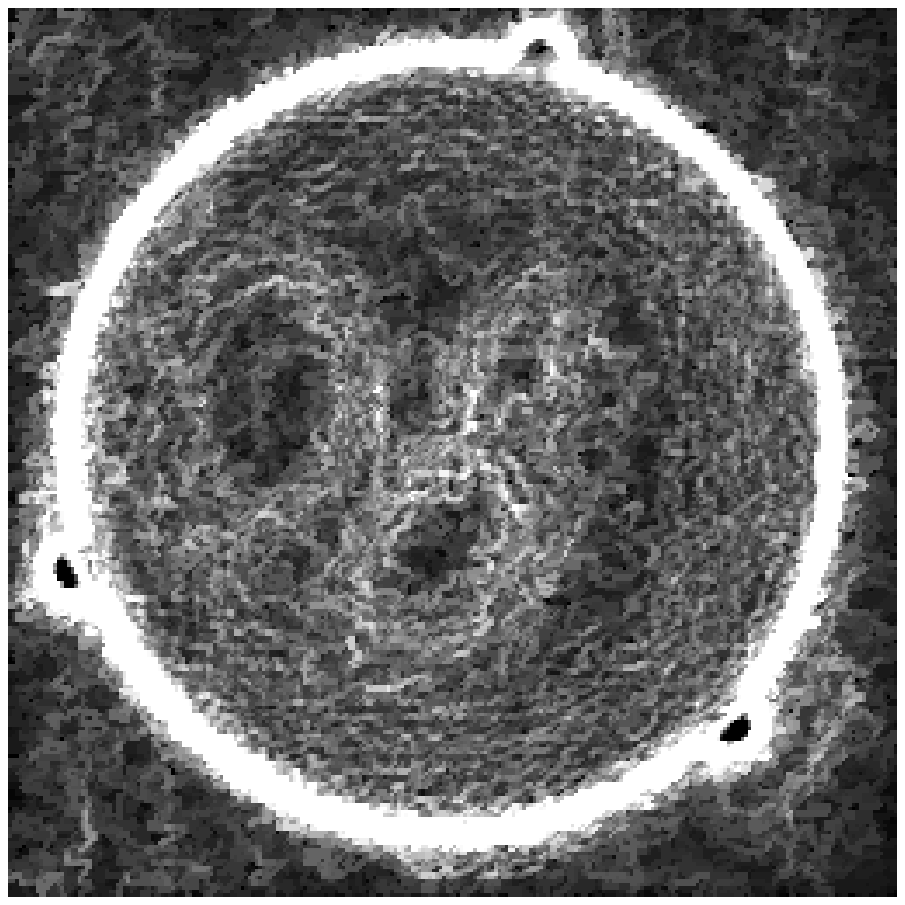}&
  \includegraphics[width=0.3\columnwidth]{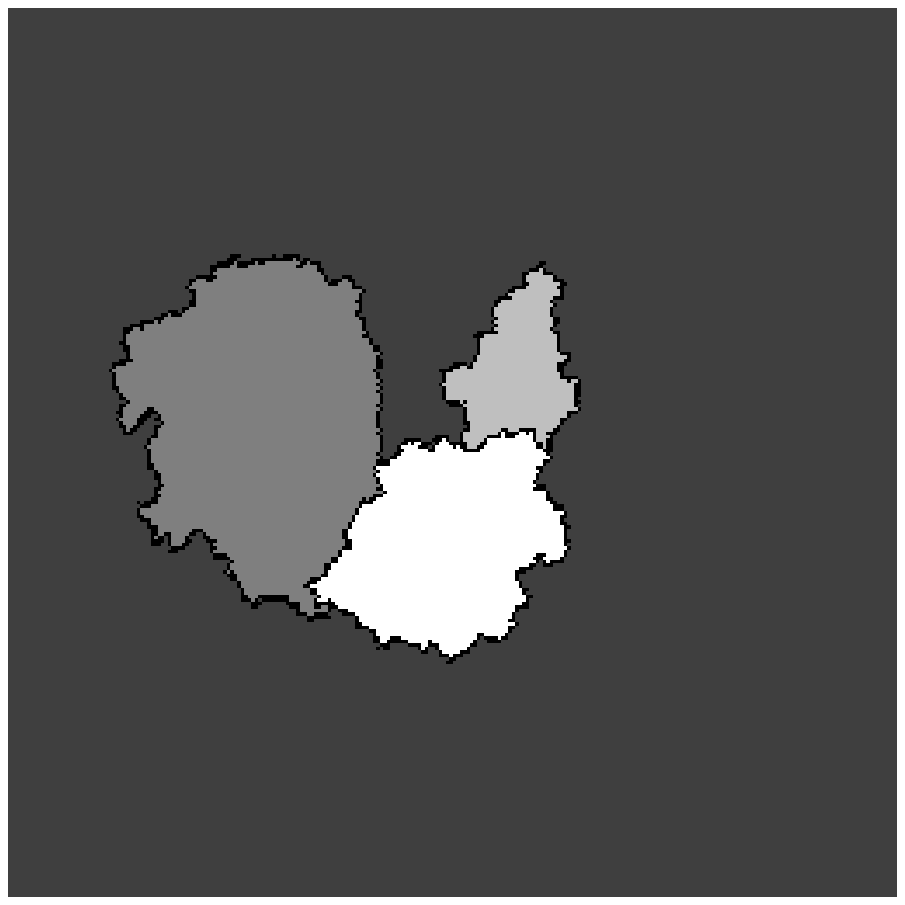}\\
  (c) & (d) \\
 \end{tabular}
\end{center}
  \caption{Segmentations on parameters: (a) Mahalanobis distance based gradient, (b) associated
  watershed, (c) gradients supremum, (d) associated watershed. For visualization, gradients are
  scaled by a factor.}
  \label{Fig_segmentation_gradient_euclidien_et_sup_para}
\end{figure}

Both segmentations from a vectorial gradient are good, compared to
the reference. Consequently vectorial gradients are good in both
cases.

\subsubsection{Watershed on PCA axes of parameters}

As for parameters, a morphological gradient is computed on each
channel of $\mathbf{c}^{\mathbf{p}}_{\beta}(x)$. The markers (fig.
\ref{Fig_marqueurs_clara_para_ACP_para} (d)) are again regularized
with an opening followed by the corresponding watershed of each
gradient.

The resulting segmentations on the parameters factors on axes 1,
$c^{\mathbf{p}}_{\beta_{1}}$, and 2, $c^{\mathbf{p}}_{\beta_{2}}$,
are good, as compared to the reference (fig.
\ref{Fig_segmentation_gradient_morpho_ACP_para}). In fact their
gradients have distinct contours without leaks. However, it is not
the case for morphological gradient on the parameters factors on
axes 3, $c^{\mathbf{p}}_{\beta_{3}}$. Moreover, as for parameters,
vectorial gradients are also tested.

First of all, an Euclidean gradient
$\nabla_{E}\mathbf{c}^{\mathbf{p}}_{\beta}(x)$ is tested. Then the
supremum of channels gradient
$\nabla_{\vee}\mathbf{c}^{\mathbf{p}}_{\beta}(x)$ and the weighted
sum gradient $\nabla_{+}\mathbf{c}^{\mathbf{p}}_{\beta}(x)$ are used
(fig. \ref{Fig_segmentation_gradient_euclidien_et_sup_ACP_para}).
The weights for $\nabla_{+}\mathbf{c}^{\mathbf{p}}_{\beta}(x)$ are
equal to the inertia contributions of axes: $0.97$, $0.021$ and
$0.0066$. For each case the same markers are used.

\begin{figure}
\begin{center}
\begin{tabular}{ccc}
  \includegraphics[width=0.3\columnwidth]{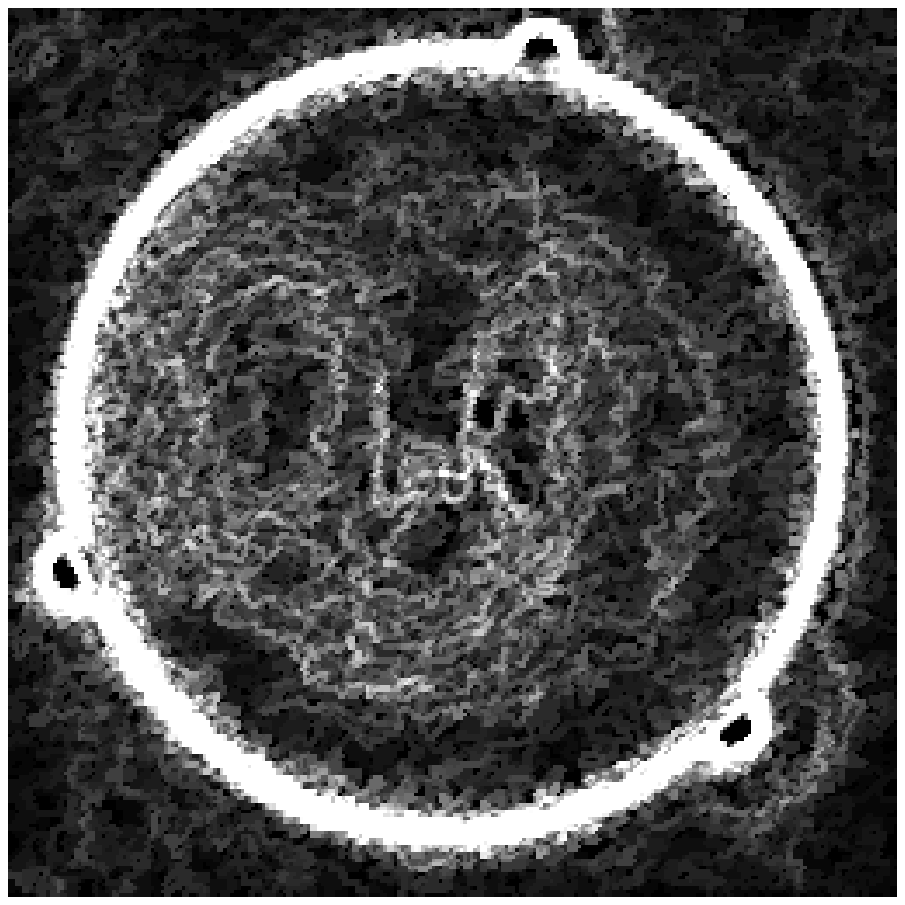}&
  \includegraphics[width=0.3\columnwidth]{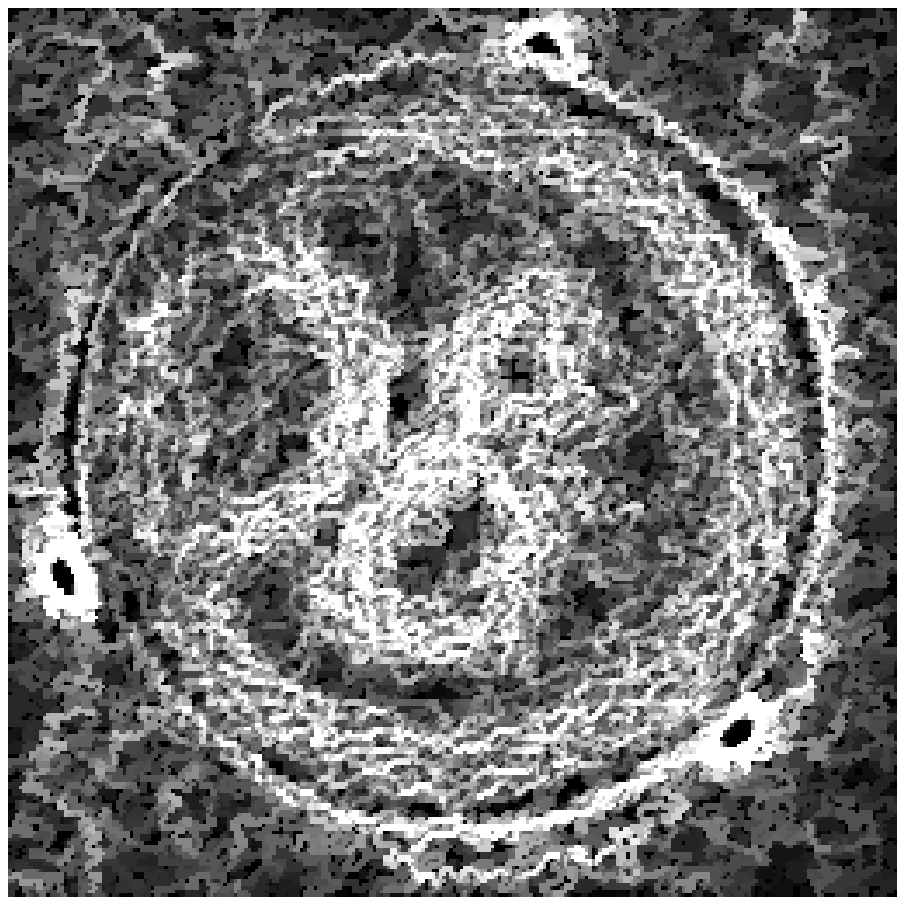}&
  \includegraphics[width=0.3\columnwidth]{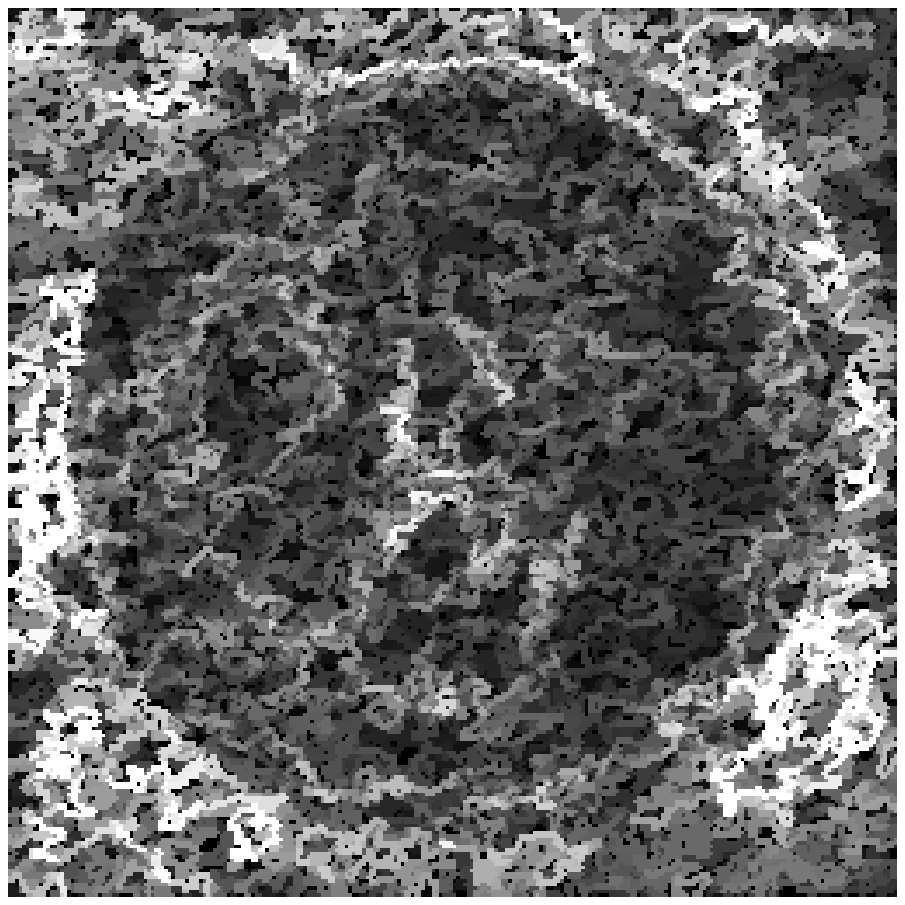}\\
  \includegraphics[width=0.3\columnwidth]{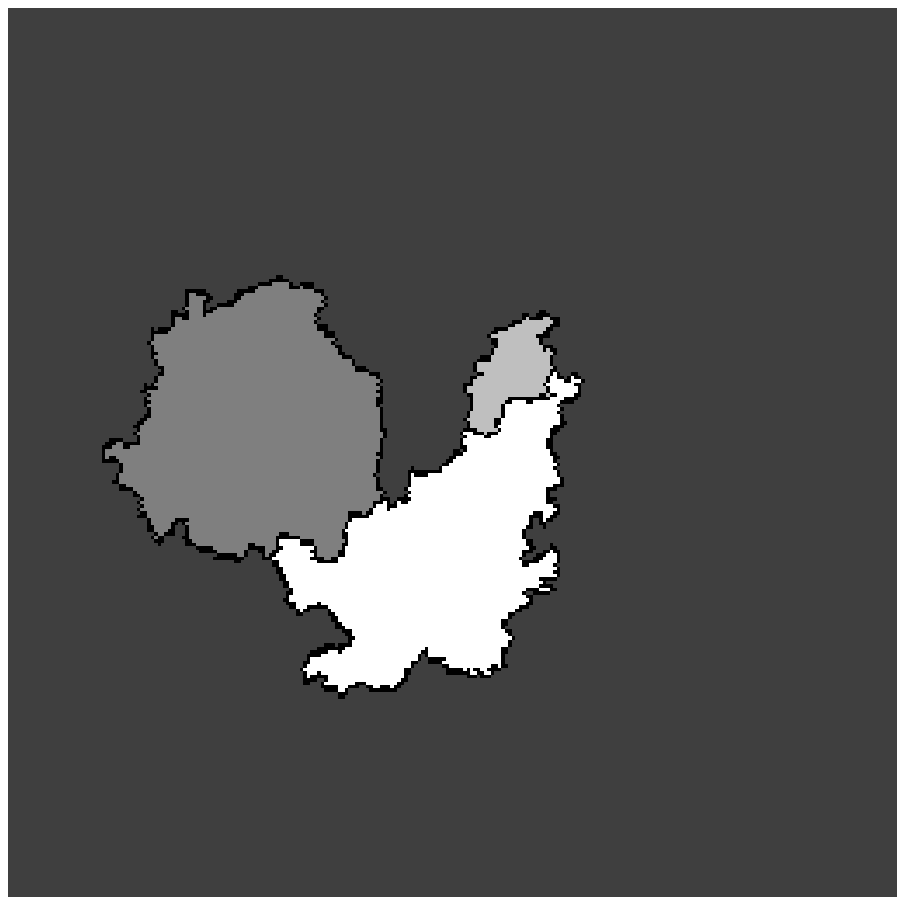}&
  \includegraphics[width=0.3\columnwidth]{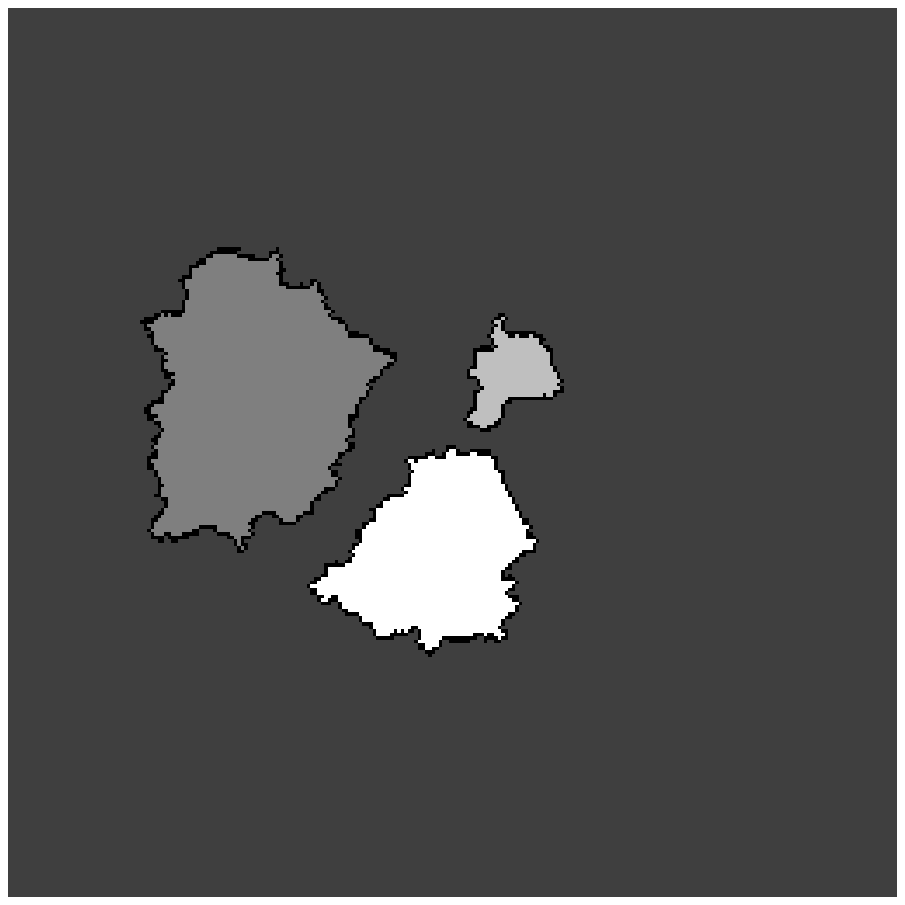}&
  \includegraphics[width=0.3\columnwidth]{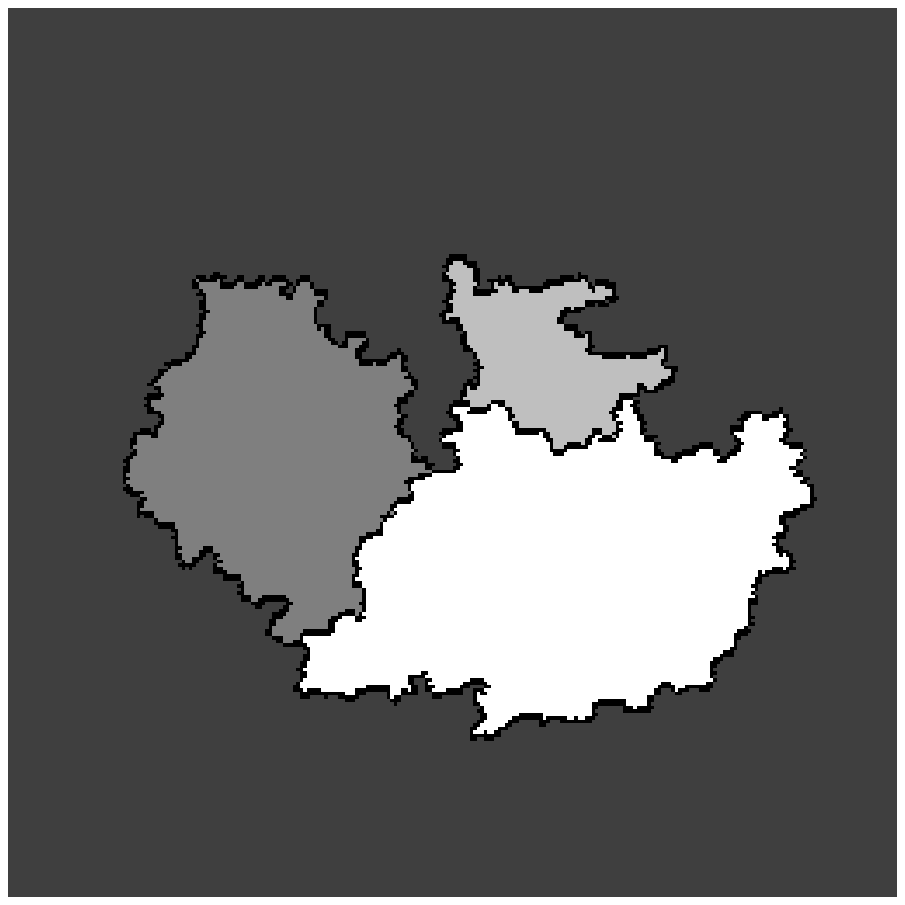}\\
  axis 1 & axis 2 & axis 3\\
\end{tabular}
\end{center}
  \caption{Morphological gradient and watershed on factors of PCA parameters: axis 1 $c^{\mathbf{p}}_{\beta_{1}}$,
  axis 2 $c^{\mathbf{p}}_{\beta_{2}}$ and axis 3 $c^{\mathbf{p}}_{\beta_{3}}$.
  For visualization, gradients are scaled by a factor. }
  \label{Fig_segmentation_gradient_morpho_ACP_para}
\end{figure}

\begin{figure}
\begin{center}
\begin{tabular}{cc}
    \includegraphics[width=0.3\columnwidth]{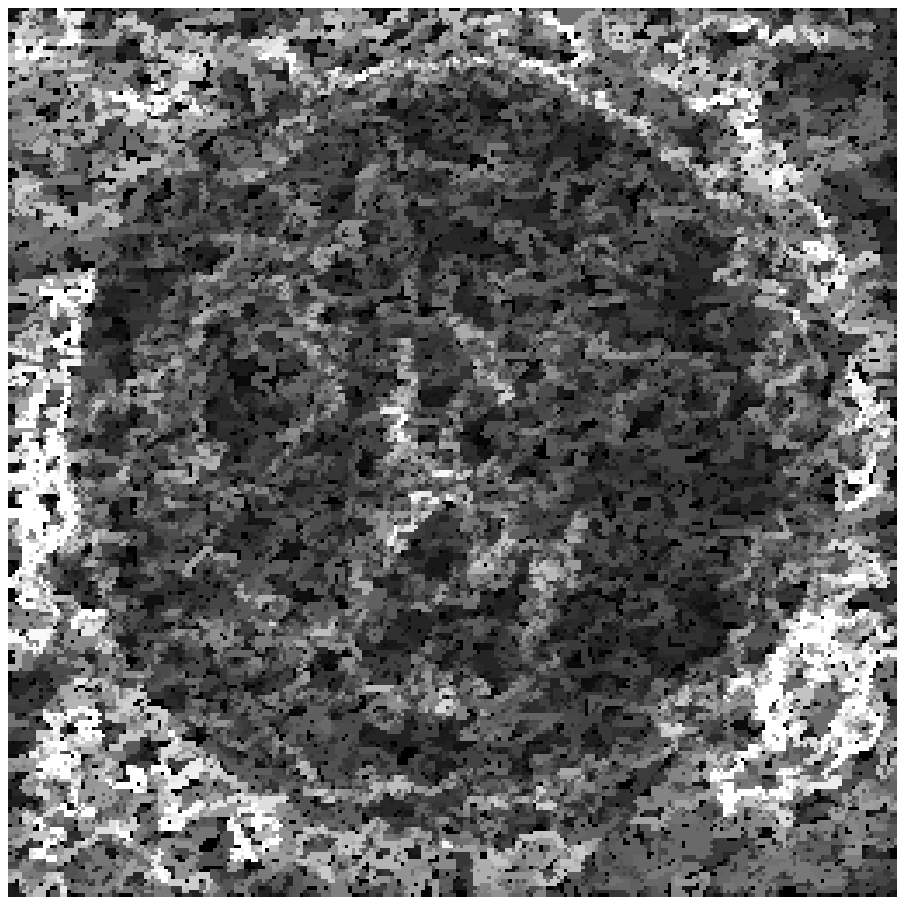}&
    \includegraphics[width=0.3\columnwidth]{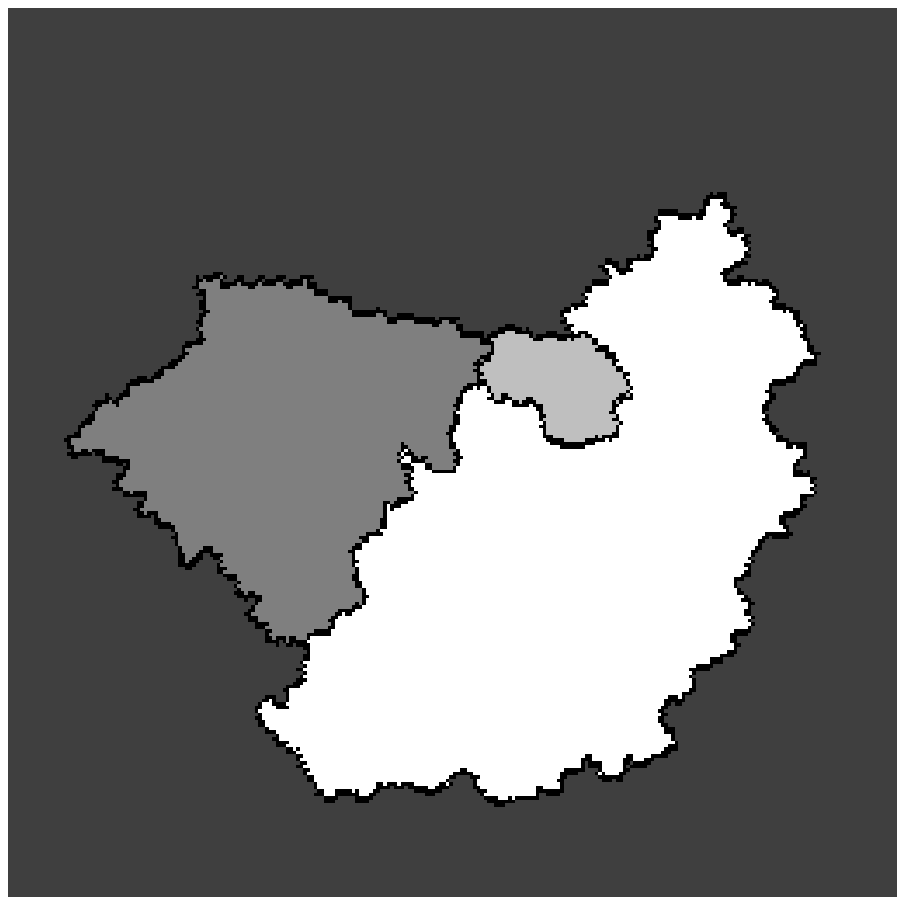}\\
    (a) & (b)\\
    \includegraphics[width=0.3\columnwidth]{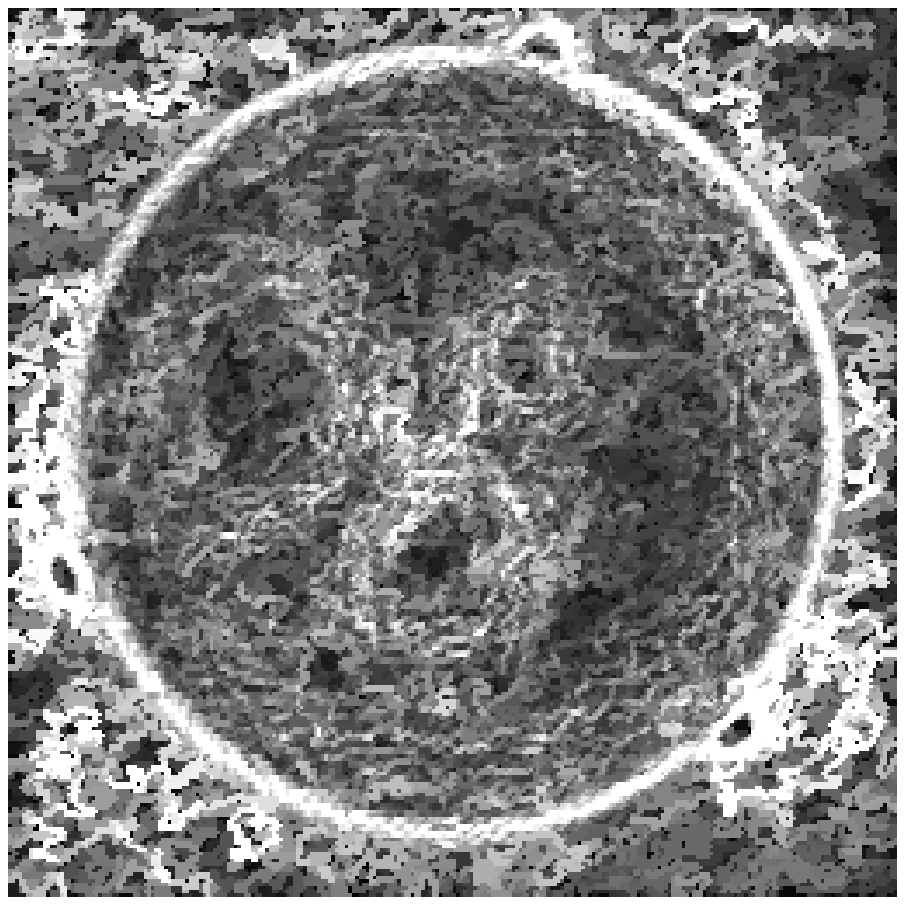}&
    \includegraphics[width=0.3\columnwidth]{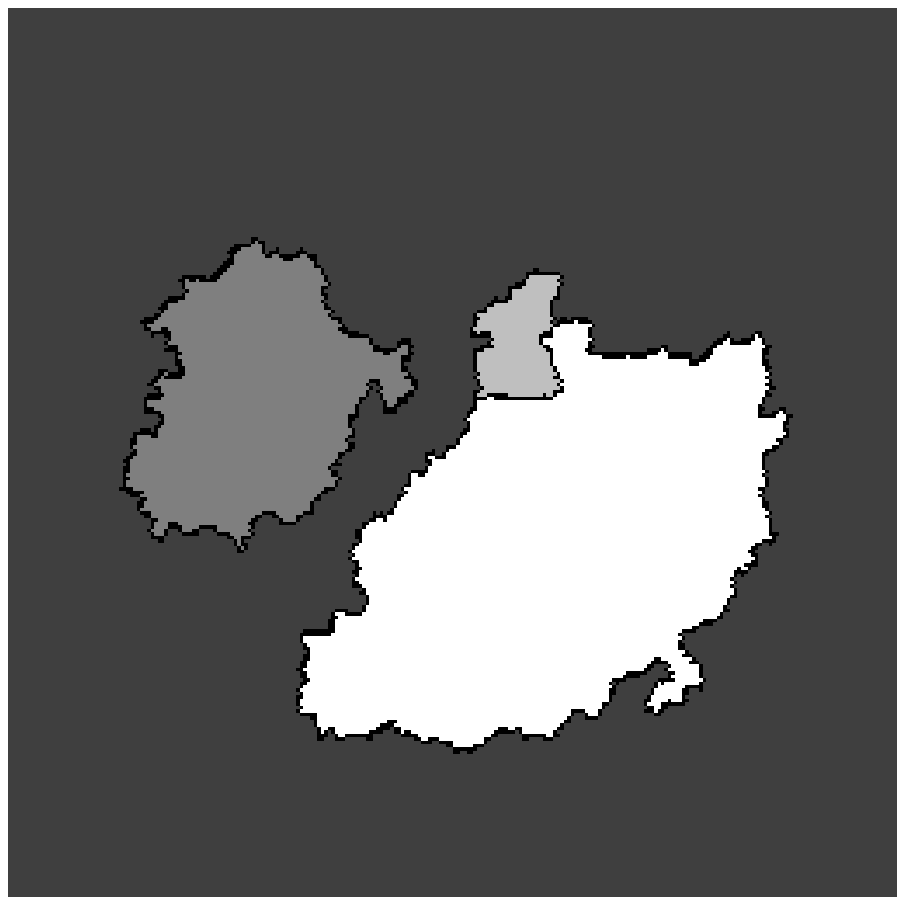}\\
    (c) & (d)\\
    \includegraphics[width=0.3\columnwidth]{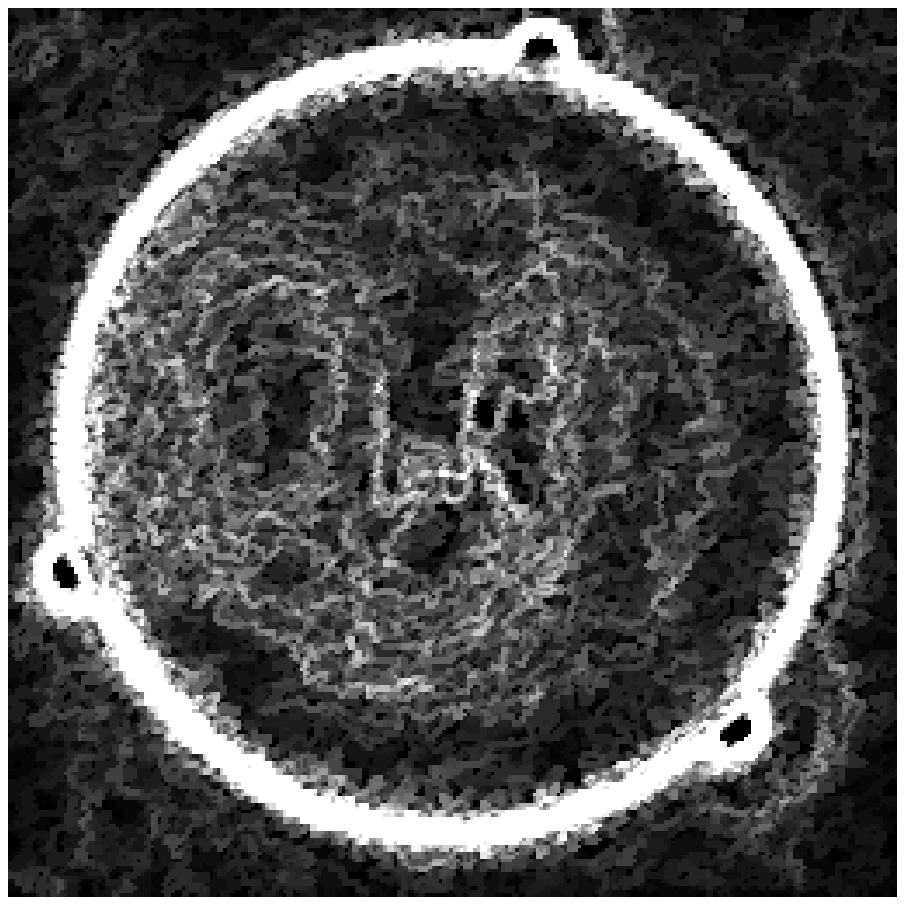}&
    \includegraphics[width=0.3\columnwidth]{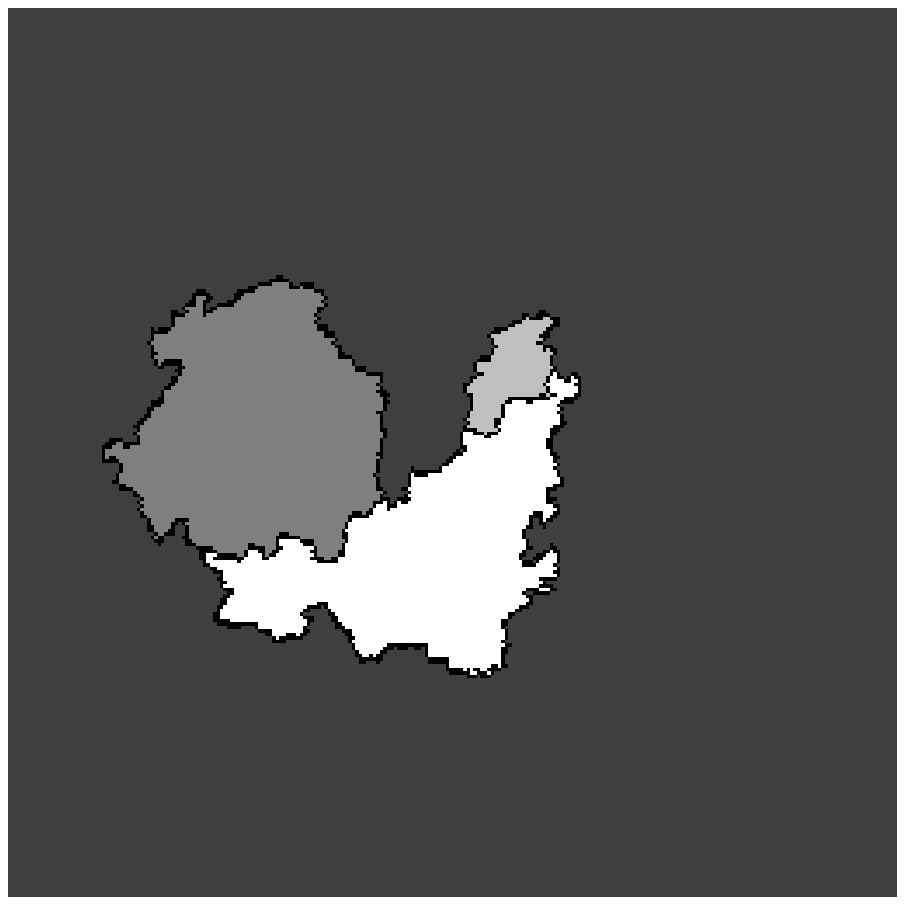}\\
    (e) & (f)\\
\end{tabular}
\end{center}
  \caption{Segmentations on PCA parameters factors on the 3 axes: (a) Euclidean gradient, (b) associated watershed,
  (c) gradient supremum, (d) associated watershed, (e) weighted sum gradient,
  (f) associated watershed. For visualization, gradients are scaled by a factor.}
  \label{Fig_segmentation_gradient_euclidien_et_sup_ACP_para}
\end{figure}

Both segmentations (fig.
\ref{Fig_segmentation_gradient_euclidien_et_sup_ACP_para}), with the
gradient supremum and the Euclidean gradient, are not good, compared
to the reference. This is due to the leaks on the gradients for the
glue occlusions. In fact the morphological gradient on the factors
on the third axis of PCA parameters is not relevant (fig.
\ref{Fig_segmentation_gradient_morpho_ACP_para}). However, the
weighted sum gradient is much better because the weight the third
axis is very low compared to the other ones. In this case the
weighted sum gradient is more adapted.

Consequently, the Euclidean gradient and the supremum of
morphological gradients are tested on the two first PCA parameters
axes (fig.
\ref{Fig_segmentation_gradient_euclidien_et_sup_ACP_para_2_axes}).
In this case, the segmentations are much better. Therefore, the
choice of axes for which the gradient is better must be emphasized.
In fact, adding another noisy axis creates leaks on the gradient.

\begin{figure}
\begin{center}
\begin{tabular}{cc}
  \includegraphics[width=0.3\columnwidth]{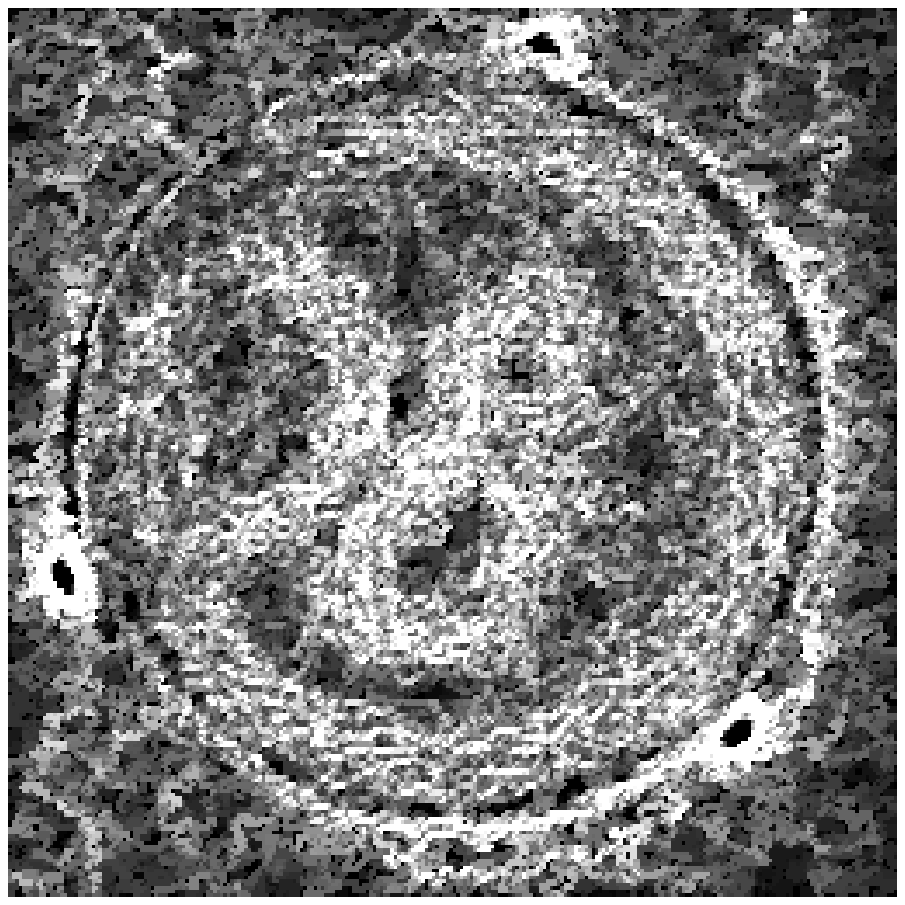}&
  \includegraphics[width=0.3\columnwidth]{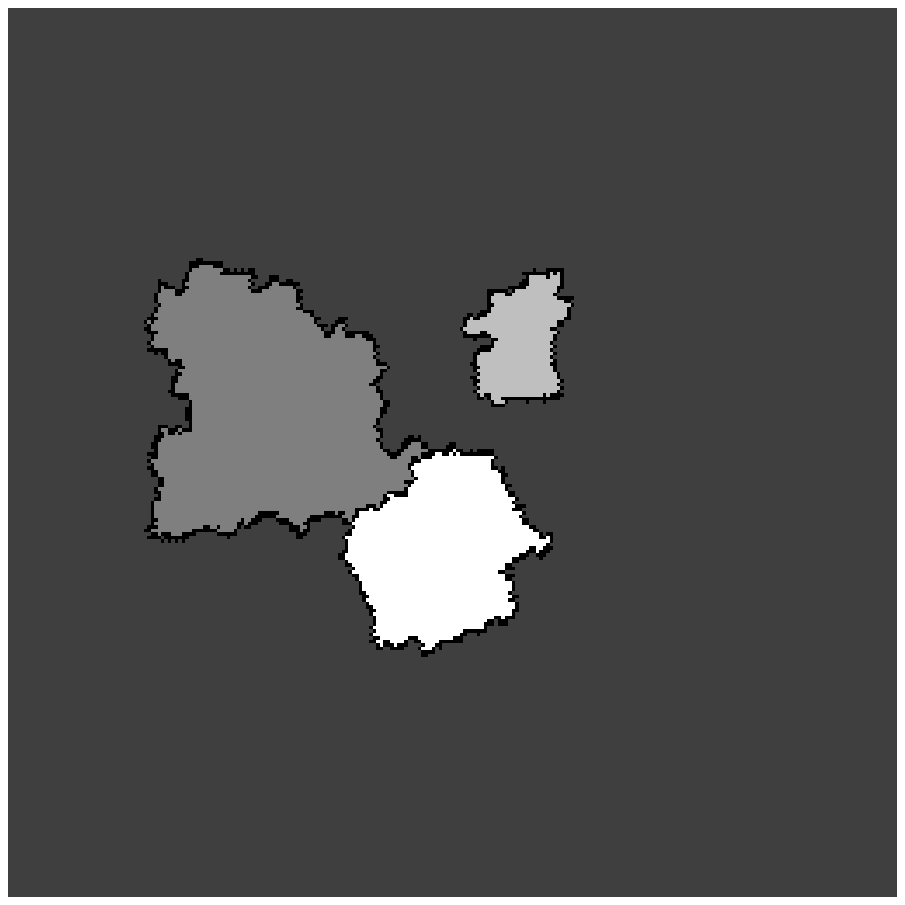}\\
  (a) & (b)\\
  \includegraphics[width=0.3\columnwidth]{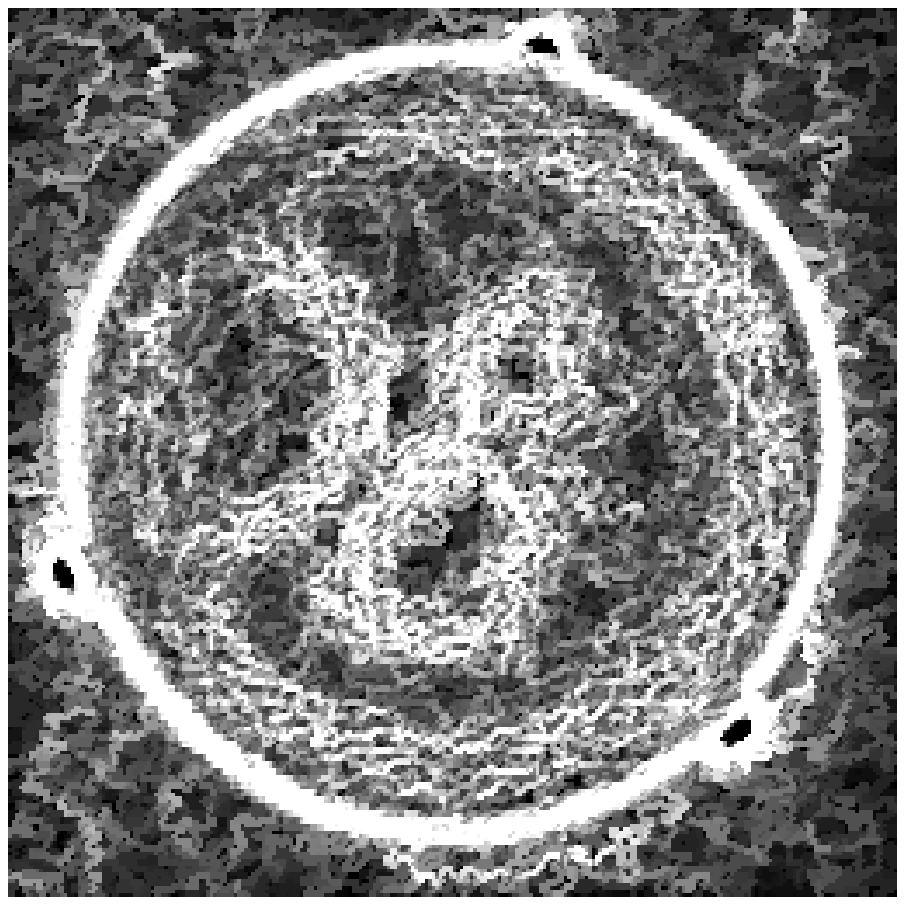}&
  \includegraphics[width=0.3\columnwidth]{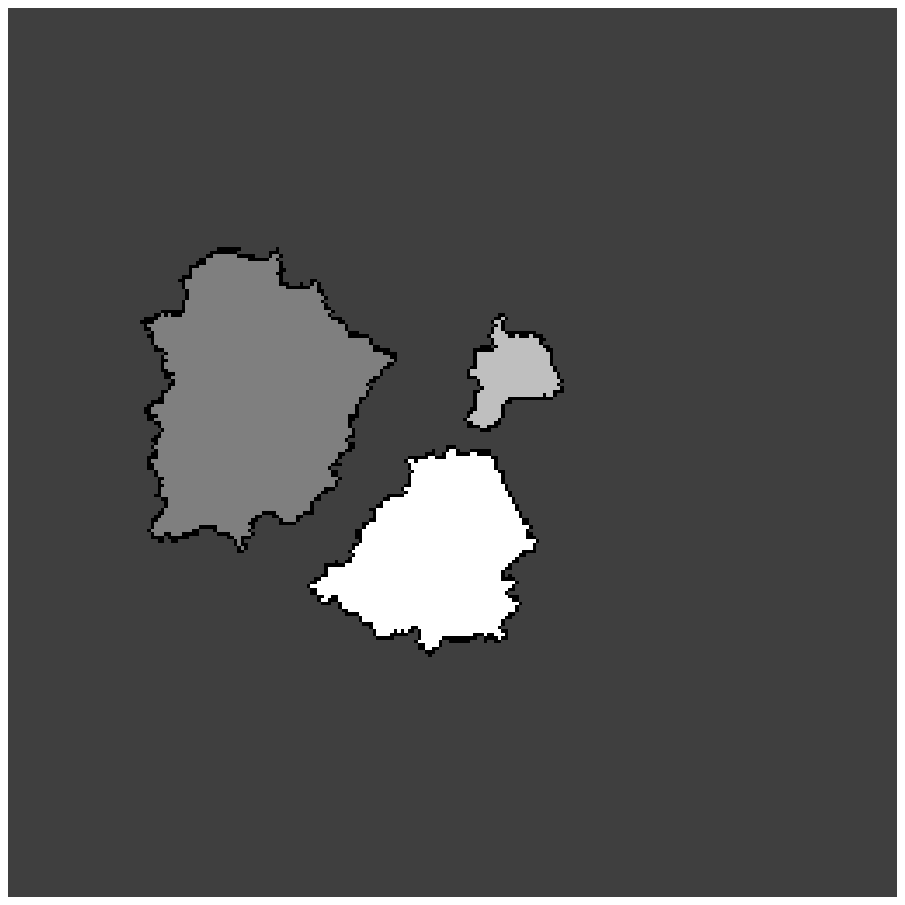}\\
  (c) & (d)\\
\end{tabular}
\end{center}
  \caption{Segmentations on PCA parameters factors on axes 1 and 2: (a) Euclidean gradient,
  (b) associated watershed, (c) gradient supremum, (d) associated watershed.
  For visualization, gradients are scaled.}
  \label{Fig_segmentation_gradient_euclidien_et_sup_ACP_para_2_axes}
\end{figure}

\section{Overview of methods}

\begin{figure}
\begin{center}
\begin{tabular}{c}
     \includegraphics[width=1\columnwidth]{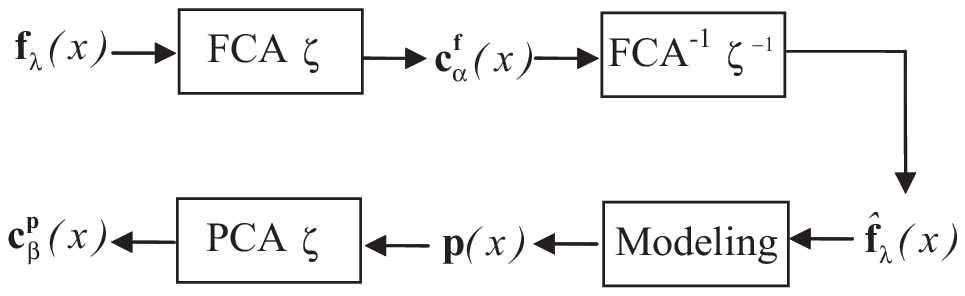}\\
     (a)\\
     \includegraphics[width=1\columnwidth]{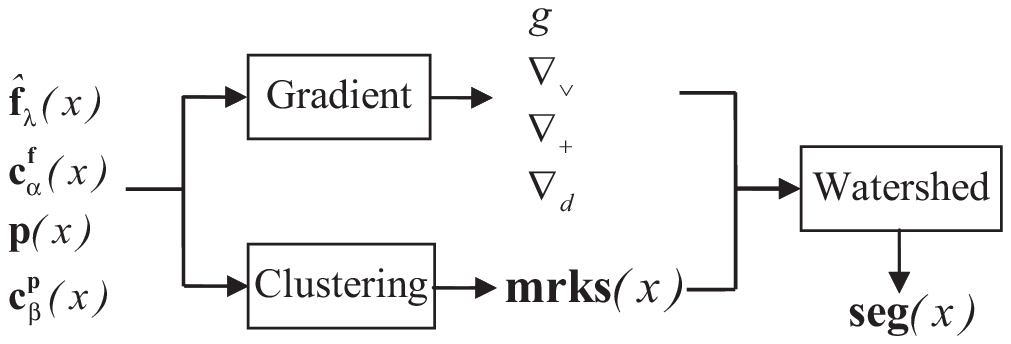}\\
     (b)\\
\end{tabular}
\end{center}
  \caption{(a) Flowchart of different transformations: data denoising and reduction,
  (b) diagram for Watershed segmentation.}
  \label{Fig_schema_process}
\end{figure}

In this paper we have considered four multidimensional spaces for image segmentation (fig.
\ref{Fig_schema_process}):
\begin{description}
  \item[$\bullet$ space 1: ] the image space $\mathbf{\widehat{f}}_{\mathbf{\lambda}}(x)$ after image filtering by FCA;
  \item[$\bullet$ space 2: ] the factorial space of image
  $\mathbf{c}^{\mathbf{f}}_{\alpha}(x)$ after image filtering and data reduction using
  FCA;
  \item[$\bullet$ space 3: ] the parameters space $\mathbf{p}(x)$ after image filtering by FCA and data
  reduction by model fitting;
  \item[$\bullet$ space 4: ] the factors parameters space
  $\mathbf{c}^{\mathbf{p}}_{\beta}(x)$ after image filtering by FCA, data reduction by model
  fitting and parameters orthogonalisation by PCA.
\end{description}

These spaces can be grouped in two different approaches. The first
one is data reduction by FCA. Space 2 belongs to this approach. The
second approach reduces data by model fitting. Spaces 3 and 4
belongs to this approach. Space 1 provides a direct approach to be
compared to the others.

On each space, the same method of segmentation can be applied. First
of all, a filtering is done on image
$\mathbf{f}_{\mathbf{\lambda}}(x)$ using FCA. Then the components of
the spaces are generated:
$\mathbf{\widehat{f}}_{\mathbf{\lambda}}(x)$,
$\mathbf{c}^{\mathbf{f}}_{\alpha}(x)$, $\mathbf{p}(x)$,
$\mathbf{c}^{\mathbf{p}}_{\beta}(x)$. These components generate new
hyperspectral images corresponding to the components space. In each
space, the same method is applied on hyperspectral images. The
segmentation combines a spectral and a spatial part. The spectral
part consists of a classification in the considered space to obtain
the markers. The spatial part consists in computing a gradient on
hyperspectral images (fig. \ref{Fig_schema_process}). Then, with the
markers of the spectral part and the gradient of the spatial part, a
watershed segmentation is performed on the considered space of
hyperspectral images.

\section{Conclusion and Perspec- \newline tives}

This paper has presented a watershed-based segmentation for
hyperspectral images. This approach combines a spectral part (the
markers) and a spatial part (the gradient).

A temporal series example is used to illustrate our methodology.
Comparing the results obtained for the various segmentations and the
reference of \citet{legrand:2002}, it is obvious that a data
reduction approach is necessary. For the current image, the data
reduction based on a model is better than the one based on factor
correspondence analysis. In fact, for an hyperspectral image, it is
better to use a model, when it can be fitted, because it reduces the
image entropy while it keeps the inner data structure. Besides, the
choice of pertinent parameters with low noise contribution is
crucial for segmentation quality.

Moreover, multivariate gradients behave better than any marginal
gradient on parameters. The multivariate gradients are based on an
adapted distance to the considered space and on the supremum, or
weighted sum, of morphological gradients on channels. The two kinds
of gradients give similar results. Besides, to obtain relevant
segmentations, they must be applied on a space with a low level of
noise. This underlines the importance of a pertinent choice for
parameters factors. About the markers, the corresponding spaces to
compute them must be also with a low level of noise, to get
pertinent ones.

In conclusion, a relevant multivariate segmentation requires an
adapted data reduction, which gives parameters or factors with a low
level of noise, is crucial ; a necessary condition to get pertinent
markers and gradients.

In the future, we will develop multivariate filtering on spectral
bands. More precisely, we will focus on the levelings. They are
usually necessary to enhance the functions on which are computed the
gradients. As for greyscale images, we will define new types of
multivariate levelings. They will be adapted to peculiarities of
these functions and they will also simultaneously filter all the
spectral channels of hyperspectral images.

In the present example we have only used a simple approach for
markers extraction, i.e. a clustering by "Clara". It is necessary to
test other classification methods combining spectral and spatial
information in order to improve markers detection. We are
considering to do clustering with lambda flat zones and clustering
after a principal coordinates analysis, using a weight/distance
matrix \citep{benzecri:1973,gower:1966}.

%
%

\bibliography{refs}
\end{paper}
\end{document}